\documentclass[11pt, floatsintext]{article}

\usepackage[english]{babel}
\usepackage[utf8x]{inputenc}
\usepackage[T1]{fontenc}
\usepackage{natbib}
\setcitestyle{authoryear, open={(},close={)}}
\usepackage{fullpage}

\usepackage{authblk}

\usepackage{amsthm,amsmath,bm,bbm}
\usepackage{amssymb,mathtools}
\usepackage{graphicx}
\usepackage[usenames,dvipsnames,svgnames]{xcolor}
\usepackage[colorinlistoftodos]{todonotes}
\usepackage{setspace}
\usepackage[font=singlespacing]{caption}
\usepackage{multicol,multirow}
\usepackage{float}

\usepackage[colorlinks=true, allcolors=blue]{hyperref}

\usepackage{cancel}
\usepackage{soul}
\usepackage{etoolbox}
\usepackage{enumitem}
\patchcmd{\maketitle}{\newpage}{}{}{}

\usepackage{tikz}
\usetikzlibrary{shapes,shadows,arrows,decorations.pathreplacing}

\AtBeginDocument{%
 \abovedisplayskip=3pt plus 2pt minus 2pt
 \abovedisplayshortskip=0pt plus 2pt
 \belowdisplayskip=3pt plus 2pt minus 2pt
 \belowdisplayshortskip=3pt plus 2pt minus 2pt
}

\def\independent{\perp\!\!\!\perp}
\def\E{\mathrm{E}}
\def\P{\mathrm{P}}

\DeclareSymbolFont{rsfs}{U}{rsfs}{m}{n}
\DeclareSymbolFontAlphabet{\mathscrsfs}{rsfs}

\def\M{\mathcal{M}}

\theoremstyle{plain}
\newtheorem*{theorem*}{Theorem}

\title{Clarifying causal mediation analysis: Effect identification via three assumptions and five potential outcomes}

\author{Trang Quynh Nguyen$^*$, Ian Schmid, Elizabeth L. Ogburn, Elizabeth A. Stuart}
\affil{Johns Hopkins Bloomberg School of Public Health}
\affil{\small $^*$email: trang.nguyen@jhu.edu}

\begin{document}

\maketitle

\begin{abstract}
    Causal mediation analysis is complicated with multiple effect definitions that require different sets of assumptions for identification. This paper provides a systematic explanation of such assumptions. We define five potential outcome types whose means are involved in various effect definitions. We tackle their mean/distribution's identification, starting with the one that requires the weakest assumptions and gradually building up to the one that requires the strongest assumptions. This presentation shows clearly why an assumption is required for one estimand and not another, and provides a succinct table from which an applied researcher could pick out the assumptions required for identifying the causal effects they target. Using a running example, the paper illustrates the assembling and consideration of identifying assumptions for a range of causal contrasts. For several that are commonly encountered in the literature, this exercise clarifies that identification requires weaker assumptions than those often stated in the literature. This attention to the details also draws attention to the differences in the positivity assumption for different estimands, with practical implications.  Clarity on the identifying assumptions of these various estimands will help researchers conduct appropriate mediation analyses and interpret the results with appropriate caution given the plausibility of the assumptions.
    
    ~
    
\end{abstract}

% \pagenumbering{gobble}
% \maketitle
% \pagenumbering{arabic}

\allowdisplaybreaks

\section{Introduction}

\noindent Causal inference analyses, explicitly or implicitly, generally involve three steps: define the target causal effect (also known as the \textit{estimand}, i.e., what we wish to estimate); assess its identifiability (what assumptions are required to learn this causal effect from observed data, and whether they are likely to hold); and then estimate it (i.e., learn it from data) \citep{Pearl2018}. If there is concern that an identification assumption may not hold, this issue should be dealt with if the analysis were to proceed, e.g., via adding a sensitivity analysis as a fourth step after estimation, or building the uncertainty about the assumption into the estimation procedure.
% \footnote{A possibility for doing this is using priors that reflect such uncertainty in a Bayesian model.}
While assumptions are part of most statistical analyses, they are especially important when inferring causal effects from observational data, because some assumptions are untestable, and if they do not hold effects may even be not interpretable. It is important that the researcher conducting an analysis understand the assumptions and judge their plausibility.
This paper clarifies for applied researchers the identifying assumptions often invoked in a \textit{causal mediation analysis} -- using a simple setting with a binary exposure $A$, a single mediator $M$ and a single outcome $Y$, with appropriate temporal ordering. Even here things are more complicated than in the non-mediation situation. Our goal is to unpack the complicated in a way that is digestible and thus helpful for practice.

Two comments before we proceed. First, this paper focuses on the single mediator case, where the mediator may be univariate or multivariate (but considered en bloc). This leaves quite a few cases outside the scope of the paper, such as settings with multiple mediators where the effects through each mediator are of interest \citep{VanderWeele2014a,Daniel2015,Vansteelandt2017} and settings with repeated exposure and mediator over a longitudinal process \citep{Zheng2017,VanderWeele2017}. 
Second, as mediation analysis concerns causal effects, exposure-mediator-ourcome temporal ordering is required. Unfortunately, reviews \citep{vo2020ConductReportingMediation,stuart2021AssumptionsNotOften} continue to find many mediation analyses not satisfying this minimal requirement, making it all the more important to reiterate. Without appropriate temporal ordering, our effect identification exercise here would be nonsense.

\subsection{Estimands}

\noindent In this paper we talk about estimands using the language of potential outcomes \citep{Rubin1974} and potential mediator values. There are a variety of causal estimands to choose from% (which, as we shall see, call for an interesting collection of identifying assumptions)
, each being a contrast of potential outcomes under two conditions. The estimands addressed in this paper are defined by conditions where the exposure and in most cases the mediator are (hypothetically) manipulated -- an approach championed by \cite{Pearl2001}.%
\footnote{Separate from the common approach of defining causal effects based on conditions where the mediator is manipulated, \cite{Robins2020} take a different approach anchored on the idea of splitting exposure into components that affect the outcome through different pathways. They define estimands as effects of  the  split  exposures,  and  develop  a  theory  with  graphical  rules  for  identification.  That  theory  is  relevant  to the research questions addressed in our illustrative example in several ways: the notion of split exposure effects speaks directly to the trimmed intervention question; some insights from the theory are indirectly helpful for the consideration of other interventional effects; and a result from the theory sheds additional light on natural (in)direct effects identification. The same research questions from the current illustrative example here could alternatively be analyzed using this split exposure theory plus a transportability lens.}
These include well-known direct and indirect effects of several types, and a range of effects that do not fit a direct or indirect effect label. We give a brief introduction of these estimands here, and return to each one when discussing identification. As the current focus is on identifying (not defining) effects, we refer the reader to the companion article \cite{Nguyen2020} for a detailed discussion of the meaning and relevance of these effect types.

\textit{Direct effects} reflect the notion of the exposure's influence on the outcome that does not go through the mediator. They are each defined based on some manner of ``blocking'' the influence that goes through the mediator. With \textit{controlled direct effects} \citep{Robins1992,Pearl2001}, this blocking is done by fixing the mediator to one value (not letting it change in response to the exposure), so a controlled direct effect is the effect of the exposure on the outcome when the mediator is fixed, and it depends on the mediator control value. With \textit{natural direct effects} \citep{Robins1992,Pearl2001}, the blocking is instead done by holding the mediator at the individual's own potential mediator value under one exposure condition. While there are as many controlled direct effects as there are possible mediator values, there are only two natural direct effects, depending on which of the two potential mediator values (under exposure and nonexposure) is used. For the \textit{interventional direct effects} \citep{Didelez2006,VanderWeele2014a}, it is the mediator distribution (rather than the individual's mediator value) that is fixed, and it is fixed to be the same as a potential mediator distribution conditional on covariates. For a set of covariates, there is one potential mediator distribution for exposure and another for nonexposure, so there are two interventional direct effects. The \textit{generalized direct effects} \citep{Didelez2006,Geneletti2007,Nguyen2020} generalize the different types of direct effects by letting the mediator distribution be held at any relevant distribution, not just a value or a potential mediator distribution.

\textit{Indirect effects} reflect the notion of the exposure's influence on the outcome that goes through the mediator. An indirect effect is defined as the effect on the outcome of a switch in the mediator value or distribution from the potential value/distribution under nonexposure to that under exposure (as if in response to a switch in the exposure), while keeping exposure unchanged. When the switch involves the individual specific potential mediator values, we have \textit{natural indirect effects}; when the switch involves the potential mediator distributions (conditional on covariates), we have \textit{interventional indirect effects}. 
Each natural indirect effect pairs with a natural direct effect in summing up to the total causal effect; in fact the motivation for the original establishment of natural (in)direct effects is to split the total effect into path-specific components.
Interventional (in)direct effects, in contrast, are not made to decompose the total effect. There are no controlled indirect effects.

All the effects above, except the natural (in)direct effects, are part of a class of effects we call \textit{interventional effects}. An effect in this class is a contrast between an \textit{interventional condition} where the exposure and/or mediator are manipulated and a \textit{comparison condition} with a different manipulation (or no manipulation); the sort of manipulation referenced here is one that sets the variable or its distribution to a value/distribution that is known or can be determined. This is a broad class that contains many effects that do not fit the notion of direct or indirect effects. An example is where the exposure is an existing intervention program, but researchers are also interested in the effect of a hypothetically modified intervention program that no longer targets a mediator, relative to the no intervention condition. For more examples, see \cite{Nguyen2020}.
% For more examples, see \cite{Nguyen2020}.
% \cite{Nguyen2020} give other examples of such effects, which to certain research questions are more relevant than the (in)direct effects.

For simplicity, causal effects are represented in this paper on the additive scale and in average form -- as differences in potential outcome mean between contrasted conditions. Alternatively, other effect scales (e.g., ratio of means) could be used, and other features of the potential outcome distribution (e.g., median) could be contrasted; the same identification assumptions apply.

% The above causal effects are defined based on conditions where the mediator (value or distribution) is manipulated. A different framework \citep{Robins2020} is based instead on the idea of splitting exposure into components that affect outcome through different causal pathways, so effects are defined as contrasts of conditions that turn on or off different exposure components (rather than intervening on the mediator). While this is not our focus, we will present our current thoughts about how ideas from this new framework may be relevant to the effects we tackle.

\subsection{Nonparametric point identification}

\noindent The type of identification discussed here -- the type commonly encountered in the causal mediation literature -- is \textit{nonparametric point identification}. Let us clarify what this means.

Since each of our estimands (an average causal effect) is a contrast of potential outcomes under two conditions, things would be simple if we were to observe both of those potential outcomes for each individual. Unfortunately, at most one potential outcome may be observed for each individual; this is called \textit{the fundamental problem of causal inference} \citep{Holland1986}. For each individual we observe the one actual outcome ($Y$) plus the exposure ($A$), the mediator ($M$) and perhaps some other variables including pre-exposure covariates ($C$) and other covariates that are affected by exposure ($L$). The key idea is, if certain assumptions hold, the estimand can be connected to the \textit{observed data distribution}, i.e., the distribution of $\{C,A,L,M,Y\}$. Specifically, the estimand is equated to a function of features of the observed data distribution (e.g., marginal or conditional means and densities). We then call the estimand \textit{identified}, or more precisely \textit{point identified}.%
\footnote{There are cases when a causal estimand is not point identified, but is bounded by functions of the observed data distribution \citep[see e.g.,][]{Miles2017}. Then it is called \textit{partially identified}. This topic is outside the scope of the current paper.}

As a well known example, in a perfect two-arm randomized controlled trial (RCT), the relevant identifying assumptions (discussed later) hold by design. The average total effect, i.e., the difference between the means of the two \textit{potential outcomes} (under treatment and control) is identified: it is equal to the difference in mean \textit{observed} outcome between the two RCT arms. But the RCT does not guarantee identification of the various other effects mentioned above, because the mediator is not randomized. 
% As far as those effects are concerned, the RCT is an observational study, and
For those effects, identification requires untestable assumptions that should be carefully considered by the researcher.

The identifying assumptions we make do not place any restrictions on the observed data distribution. This means no parametric assumptions such as the type of distribution (normal or other) of variables or the functional form (linear or other) for the associations among variables. The identifying assumptions we make are about equating certain conditional means or densities of potential outcomes (or potential mediators) with conditional means or densities of observed variables, the latter being free to be what they are. This type of identification is thus called \textit{nonparametric identification}.%
\footnote{There are cases where effects are nonparametrically not identified, but the researcher uses a parametric assumption as a remedy. Examples include the no exposure-mediator interaction assumption when the cross-world independence assumption required by natural (in)direct effects is violated, or the use of model extrapolation to compensate for violation of positivity. Such \textit{parametric identification} is outside the scope of the current paper.}

Note that this paper addresses identification, not estimation. Questions such as what models should be fit, or how much can be learned from a sample of a certain size, belong in the estimation step. To put them aside, it may be helpful to imagine having infinite data. 

% A note of caution: the focus of this paper is identification, not estimation. Of course, the identification results -- which link the estimand to the observed data distribution -- may appear to suggest how the estimand might be estimated. However, such apparent way is usually not the only way, and is sometimes a suboptimal way. A related point is that, in considering identification, we are not (yet) concerned about whether there is much information from a sample of a certain size, which is a question for the estimation step.

\subsection{Effect identification via five potential outcome types and three assumptions}

\noindent Much has been written about assumptions for identification of specific effects in causal mediation analysis \cite[see e.g.,][etc.]{Pearl2001,Petersen2006,Imai2010,VanderWeele2014a,Robins2020}. The current paper focuses on a systematic explanation of the assumptions, so that the logic of why an assumption is required for one estimand and not another is clear, and the reader can pick out which assumptions are required for the causal effect they target.

Causal effect identification amounts to identifying the two potential outcome means in the contrast. We organize the potential outcomes involved in the various causal effects mentioned above into five types and explain the identifying assumptions required for each, starting with the type that requires the weakest and gradually building toward the one that requires the strongest, assumptions, clarifying connections from one type to the next. 

We show that identification of the mean (or distribution) of each potential outcome requires three different types of assumptions. 
% (The different assumptions found in the literature for various effects are different versions that are specific to the potential outcome types involved.) 
We refer to them as \textit{consistency}, \textit{conditional independence} and \textit{positivity}, but note that they have been discussed under various names in the literature. \textit{Consistency} assumptions \citep{VanderWeele2009,Cole2009} are closely related to \citeauthor{Rubin1974}'s (\citeyear{Rubin1974}) \textit{stable unit treatment value assumption} (SUTVA). \textit{Positivity} is also known as \textit{overlap} or \textit{common support} in the context of identifying causal contrasts. Assumptions in our \textit{conditional independence} category have been called \textit{unconfoundedness}, \textit{ignorability}, \textit{exchangeability}, \textit{conditional randomization}, etc. \citep[see e.g.,][]{Hernan2019, Imbens2000,Imbens2008,Rosenbaum1983,VanderWeele2009}. As assumptions in this category are formally stated using conditional independence statements -- of certain variables with potential outcomes (potential mediators) -- we adopt the shorthand label \textit{conditional independence}, which allows concise reference to specific assumption components.

% We use \textit{conditional independence} as shorthand for assumptions that exposure is conditionally independent of potential outcomes (potential mediators), or that mediator (or potential mediators) is conditionally independent of potential outcomes. These have been called \textit{unconfoundedness}, \textit{ignorability}, \textit{exchangeability}, \textit{conditional randomization}, etc. \citep[see e.g.,][]{Hernan2019, Imbens2000,Imbens2008,Rosenbaum1983,VanderWeele2009}. The notorious \textit{cross-world conditional independence} assumption belongs in this category; it is relevant to the fifth potential outcome mean. 

A note for readers not familiar with the concept of conditional independence: that variables $A$ and $B$ are independent \textit{conditional on} (or \textit{given}) variable $C$, formally $A\independent B\mid C$, means that within levels of $C$ (or within each subpopulation that shares the same value on $C$), knowing $A$ does not tell us anything about $B$ and vice versa. In our current problem, the place of $B$ is occupied by a potential outcome or potential mediator; the $A$ place in most assumptions is occupied by the observed exposure or mediator, except in one assumption it is occupied by a potential mediator; and the $C$ place is occupied by covariates.

% After establishing the assumptions for each type, we apply them to examine the full set of assumptions needed to identify one or more relevant causal effects.

% , and tackle identification of their means in an order from the easiest to the hardest (i.e., requiring the weakest to the stronger assumptions) to identify. This reveals a gradual building of the assumptions and makes it meaningful. This presentation also clarifies that there are basically three assumptions that are required for identifying a potential outcome mean, and the different sets of assumptions seen in the literature are different versions of these assumptions that are specific to the potential outcome type.

% The paper proceeds as follows. The second section introduces the five potential outcome types and explains how their means are relevant to the different causal effects. The third section tackles the easiest-to-identify potential outcome mean, the mean of the potential outcome in a world where exposure is set to one condition (either exposed or unexposed). Anchoring on this potential outcome mean, this section shows how three assumptions, termed \textit{consistency}, \textit{conditional independence} and \textit{positivity}, combine to achieve identification. The fourth section builds on this foundation to explain more complex versions of these assumptions that are required for identifying the means of the other four potential outcome types, which correspond to conditions where both the exposure and the mediator are manipulated, where the variation is in the manipulation.

\subsection{Illustrative example}

% [IAN'S COMMENT: i think you need to say more up front about this instead of revealing details later on. given how prominently you feature the city program alongside the intervention in the first few illustrative examples, i think you need to talk about it up here. that said, i challenge the realism of the city program example. are there real-life examples in which a city takes responsibility for its residents with psychiatric conditions? i think it would be much more realistic if the patients were situated in some sort of health-care organization like an ACO. also, i think you need to be explicit here about (a) how long the two intervention components last, and (b) when symptom management and effective service use are measured relatively. 6 and 12 months doesn't mean anything without saying the length of the intervention components. also, you later on refer to the intervention being provided free-of-charge, but haven't mentioned that here.]

\noindent After establishing the identifying assumptions specific to each potential outcome type, we apply them to examine the full set of assumptions needed to identify one or more relevant causal effects.
We illustrate this using a running fictional example.
% This example involves intervention that is loosely based on \citeauthor{Kilbourne2008}'s (\citeyear{Kilbourne2008}) bipolar disorder medical care model.
%
In this example, of interest is the health of people who have both a psychiatric disorder and chronic medical problems, and the issue is that psychiatric symptoms may pose challenges to the patient's access to and effective use of medical care \citep{Zeber2009}. We restrict our attention to people who are members of a (more or less) organized system for the provision of health care; in the US this could be a health maintenance organization or an accountable care organization. The members all have health insurance coverage, and the system has some ability to enact certain system-wide change in the practice of care.

Suppose a local (e.g., state) branch of the system offers an intervention program that aims to improve outcomes for members with a bipolar diagnosis and a chronic medical problem, at no additional charge. (This intervention is loosely based on \citeauthor{Kilbourne2008}'s (\citeyear{Kilbourne2008}) bipolar disorder medical care model.) It consists of (i) a self-management education component that teaches patients (in group sessions over a three-month period) how to manage chronic psychiatric and medical conditions, improve dietary behavior and physical activity, and communicate better with medical providers; and (ii) a care-management component (starting at about four months) that involves help from a case manager who facilitates the patient's communication with their medical providers and oversees the patient's health care. 

With this \textit{intervention}, we take the outcome to be the patient's health-related quality of life at 18 months after program enrollment (\textit{quality of life}). Two variables are theorized to be on the causal path: a measure of proficient self management of psychiatric symptoms at three months and a measure of effective use of medical care services at 12 months. We label these variables \textit{symptom management} and \textit{service use} for conciseness, noting that the second variable is about effectiveness (not quantity) of service use. We illustrate causal contrasts of several types, some focusing on the intervention itself, others in consideration of context changes or system-wide practice adjustments.

% \subsection{A note on a different framework}

% \noindent Separate from the common approach of defining causal effects based on conditions where the mediator is manipulated, \cite{Robins2020} take a different approach anchored on the idea of splitting exposure into components that affect the outcome through different pathways. They define estimands as effects of the split exposures, and develop a theory with graphical rules for identification. That theory is relevant to the research questions addressed in our illustrative example in several ways: the notion of split exposure effects speaks directly to the trimmed intervention question; some insights from the theory are indirectly helpful for the consideration of other interventional effects; and a result from the theory sheds additional light on natural (in)direct effects identification. Our complementary work \citep{Nguyen2021applyRobins} analyzes the same research questions from the current illustrative example using this split exposure theory plus a transportability lens. 

\section{Five potential outcome types}

\subsection{First type: $Y_a$}

\noindent  $Y_a$ is the potential outcome in a world where exposure is set to $a$, where $a$ may be 1 (exposed) or 0 (unexposed). We denote its mean by $\E[Y_a]$, with the general notation $\E[\cdot]$ indicating \textit{expectation} (or \textit{mean}). This potential outcome type is relevant to the \textit{average total effect}, defined as $\text{TE}=\textcolor{blue}{\E[Y_1]}-\textcolor{blue}{\E[Y_0]}$, the difference between mean potential outcome under exposure and mean potential outcome under nonexposure. $\E[Y_a]$ is also involved in the natural (in)direct effects mentioned earlier, which decompose the total effect. (These will be formally defined shortly.) In addition, $\E[Y_a]$ is relevant to effects of all (hypothetical) intervention conditions that are assessed relative to the existing nonexposure condition.

\subsection{Second type: $Y_{am}$}

\noindent This is the potential outcome in the hypothetical world where exposure is set to $a$ and mediator is set to a specific value $m$. $Y_{am}$ is relevant to controlled direct effects, as formally the controlled direct effect for a mediator control level $m$ is $\text{CDE}(m)=\textcolor{blue}{\E[Y_{1m}]}-\textcolor{blue}{\E[Y_{0m}]}$. In addition, $\E[Y_{am}]$ is the building block for identification of the next potential outcome means -- where we consider not just one mediator value but a range of values under a distribution.

\subsection{Third type: $Y_{a\M}$ where $\M$ is a known distribution}

\noindent This is the potential outcome in a hypothetical world where the exposure is set to $a$ and the mediator is intervened upon and set to a distribution $\M$ (which we call the \textit{interventional mediator distribution}), where $\M$ is a distribution that is either known or is defined based on data that are observed. Seen from the individual perspective, each individual is assigned a mediator value randomly drawn from the distribution $\M$. This is thus a \textit{stochastic} intervention \citep{DiazMunoz2012} on the mediator, whereas the intervention corresponding to the second potential outcome type is \textit{deterministic}.

% This may also be viewed as each individual being assigned a mediator value randomly drawn from the distribution $\M$; see \cite{Nguyen2020} on the dual population-individual view. This intervention on the mediator is a stochastic intervention \citep{DiazMunoz2012}.

$Y_{a\M}$ is relevant to generalized direct effects where the mediator distribution is fixed to a distribution $\M$, i.e., $\text{GDE}(\mathcal{M})=\textcolor{blue}{\E[Y_{1\mathcal{M}}]}-\textcolor{blue}{\E[Y_{0\mathcal{M}}]}$. In addition, it is relevant to a wide range of interventional effects where in the active intervention condition (or in both conditions contrasted) the potential outcome is of the $Y_{a\M}$ type. Applications 3 and 4 in our illustrative example (see sections \ref{app3} and \ref{app4}) concern such effects.

Depending on the specific condition of interest, the interventional mediator distribution $\M$ may be defined unconditionally (i.e., the same distribution applies to everyone) or conditional on pre-exposure covariates (e.g., different distributions for men and women) or on post-exposure covariates.
In the latter case, we require a \textit{same-world} rule: the distribution $\M$ may be defined conditional on $L_a$ (which arises after exposure has been set to $a$), but not on $L_{a'}$ (where $a'$ is the other exposure condition) or on the observed $L$ (a mixture of $L_a$ and $L_{a'}$). We can think of this as a plausible interventional world: after exposure is set to $a$, only $L_a$ arises, so an intervention on the mediator may condition on $C$ and $L_a$.

\subsection{Fourth type: $Y_{a\M}$ with $\M$ defined based on potential mediator distribution(s)}

\noindent This is similar to the third type, except that $\M$ is defined based on potential mediator distribution(s). This type is involved in interventional (in)direct effects. Recall that an interventional direct effect is the effect of exposure on outcome had the mediator distribution been fixed to be the same as a potential mediator distribution, and an interventional indirect effect is the effect on the outcome of a switch in the mediator distribution from the potential distribution under nonexposure to that under exposure while exposure itself is fixed. These are formally $\text{IDE}_{a^*}=\textcolor{blue}{\E[Y_{1\mathcal{M}_{a^*\mid C}}]}-\textcolor{blue}{\E[Y_{0\mathcal{M}_{a^*\mid C}}]}$ and $\text{IIE}_a=\textcolor{blue}{\E[Y_{a\mathcal{M}_{1\mid C}}]}-\textcolor{blue}{\E[Y_{a\mathcal{M}_{0\mid C}}]}$, where $\mathcal{M}_{a^*\mid C}$ ($a^*$ being either 0 or 1) is convenient notation indicating that the interventional mediator distribution $\M$ is defined to be the same distribution as that of the potential mediator $M_{a^*}$ given $C$.

% ($a^*$ being either 0 or 1), where $\mathcal{M}_{a^*\mid C}$ is convenient notation indicating that the interventional mediator distribution $\M$ is defined to be the same as the distribution of potential mediator $M_{a^*}$ given $C$.

% For example,  $\text{IDE}_a=\textcolor{blue}{\E[Y_{1\mathcal{M}_{a\mid C}}]}-\textcolor{blue}{\E[Y_{0\mathcal{M}_{a\mid C}}]}$, where $\mathcal{M}_{a\mid C}$ (for $a=0,1$) is convenient notation indicating that the interventional mediator distribution is defined to be the same as the distribution of $M_a$ given $C$

% e.g., $\text{IDE}_a=\textcolor{blue}{\E[Y_{1\mathcal{M}_{a\mid C}}]}-\textcolor{blue}{\E[Y_{0\mathcal{M}_{a\mid C}}]}$ and $\text{IIE}_a=\textcolor{blue}{\E[Y_{a\mathcal{M}_{1\mid C}}]}-\textcolor{blue}{\E[Y_{a\mathcal{M}_{0\mid C}}]}$,
% where $\mathcal{M}_{a\mid C}$ (for $a=0,1$) is convenient notation indicating that the interventional mediator distribution is defined to be the same as the distribution of $M_a$ given $C$ \citep{Didelez2006,VanderWeele2014a}.
Like the previous potential outcome type, this fourth type is relevant to a wide range of interventional effects, not limited to interventional (in)direct effects. Applications 5 and 6 in the illustrative example (sections \ref{app5} and \ref{app6}) concern such effects.

With this potential outcome type, we make an important subtle differentiation between the interventional mediator distribution $\M$ and the potential mediator distribution(s): the former is defined based on the latter, the latter informs the former, but the two are not the same. This allows us to consider a simple potential mediator type, $M_{a^*}$, the potential mediator if exposure were set to $a^*$ ($a^*$ being just a separate index that is not tied to $a$), but have ample flexibility in defining the distribution $\M$. For example, $\M$ could be defined based on the distribution of $M_1$ only, or of $M_0$ only (as in the interventional (in)direct effects above), or a mixture of both. $\M$ could be defined unconditionally, e.g., to be the same distribution as the marginal distribution of $M_{a^*}$, or could be defined conditional on $C$ to be the same as the distribution of $M_{a^*}$ given $C$. $\M$ could also be defined conditional on $(C,L_a)$ to be the same distribution as that of $M_{a^*}$ given $(C,L_{a^*})$. Note that in all these cases, the interventional mediator distribution $\M$ respects the same-world rule.

\subsection{Fifth type: the cross-world potential outcome $Y_{aM_{a'}}$, with $a\neq a'$}

\noindent This is the potential outcome in a completely imaginary world where the exposure is set to condition $a$ and then the mediator is set, for each individual, to its potential value under the opposite condition $a'$. There are two potential outcomes of this type, $Y_{1M_0}$ and $Y_{0M_1}$. They decompose the total effect into two pairs of natural (in)direct effects:
$$\text{TE}=\overbrace{\E[Y_1]-\textcolor{blue}{\E[Y_{1M_0}]}}^{\textstyle\text{NIE}_1}\,+\,\overbrace{\textcolor{blue}{\E[Y_{1M_0}]}-\E[Y_0]}^{\textstyle\text{NDE}_0}~~\text{and}~\text{TE}=\overbrace{\E[Y_1]-\textcolor{blue}{\E[Y_{0M_1}]}}^{\textstyle\text{NDE}_1}\,+\,\overbrace{\textcolor{blue}{\E[Y_{0M_1}]}-\E[Y_0]}^{\textstyle\text{NIE}_0}.$$

\bigskip

\noindent Additional note: The five potential outcome types above are not exhaustive. These types treat the exposure differently from the mediator: the exposure is set to one value (with no randomness), while the mediator is set either to one value or to a distribution (that is, with randomness). There may be cases in which we are interested in a condition where exposure is set to a certain distribution $\mathcal{A}$ instead of a single value 1 or 0 (e.g., where an outreach campaign helps substantially increase the rate of enrollment in the intervention program but not make it 100\%).
% This is relevant when considering interventions that shift the exposure distribution, e.g., two different outreach campaigns that result in different rates of enrollment in the bipolar care program. 
We do not consider this potential outcome type separately, because identification for the five listed types renders this type identified, e.g., if the distribution $\mathcal{A}$ is that of 1/3 exposed and 2/3 unexposed, then $\E[Y_{\mathcal{A}}]=(1/3)\E[Y_{1}]+(2/3)\E[Y_{0}]$.

\section{Identification of $\E[Y_a]$}

% \noindent Identification of the mean (or distribution) of a potential outcome of any type requires three assumptions, \textit{consistency}, \textit{conditional independence} and \textit{positivity}, each with variation depending on the specific potential outcome, which we examine one by one. These assumptions for all of the five potential outcomes are collected in Table \ref{tab:3x5}.
% In the literature these assumptions are discussed under a range of names. \textit{Conditional independence} is referred to in terms of \textit{unconfoundedness}, \textit{ignorability}, \textit{exchangeability}, (as if) \textit{randomization} \citep[see e.g.,][]{Hernan2019, Imbens2000,Imbens2008,Rosenbaum1983,VanderWeele2009}. \textit{Positivity} is also often known as \textit{overlap} or \textit{common support}. \textit{Consistency} is closely related to \citeauthor{Rubin1974}'s (\citeyear{Rubin1974}) \textit{stable unit treatment value assumption} (SUTVA). We connect to some of these terms below. 

% We now start with the specific assumptions that identify $\E[Y_a]$. Generally, \textit{consistency} is rather simple; \textit{positivity} is straightforward after \textit{conditional independence} is clear; and \textit{conditional independence} is the more complicated assumption, but for $\E[Y_a]$ it is also simple.

\subsection{Consistency assumption}

\noindent \textit{Consistency} is the type of assumption that connects potential variables to observed variables. Here the assumption is that $Y=Y_a$ if $A=a$. That is, for individuals with actual exposure $A=a$, the observed outcome $Y$ reveals the potential outcome $Y_a$. This seems like an obvious fact, but it is an assumption; the idea is that the potential outcome $Y_a$ is well defined \citep{Cole2009}, no matter how exposure value $a$ is assigned to the individual and no matter what exposure is assigned to other individuals \citep{Rubin1974}. %\citep[see also SUTVA in][]{Rubin1974}.
% \footnote{If familiar with the literature on the potential outcome framework, the reader may recognize that this coincides with a component Rubin's SUTVA, which is invoked for the definition of the potential outcome $Y_a$ for the individual.}
This assumption would be violated if one person's exposure affects others' outcomes (a typical example being vaccination), or, say, if a person's potential outcome under an exposure varies depending on whether they self-select or are assigned the exposure.

\subsection{Leveraging covariates}

\noindent \textit{Consistency} takes us one step toward our goal; it says that we observe $Y_a$ in \textit{some} individuals. But there is a missing data problem, as we want the mean of $Y_a$ over the \textit{whole} population. 
To handle this problem, the strategy is to leverage a set of observed covariates $C$ 
% (that matter to the potential outcome $Y_a$) 
such that conditional on $C$ (i.e., within each subpopulation that shares the same $C$ values), the mean of $Y_a$ is identified from the partially observed data. Consequently, the population mean is identified, as it is basically a weighted average of the subpopulation means, where the weights reflect the distribution of the covariates. This is written formally as  
$$\E[Y_a]=\E_C\{\,\E[Y_a\mid C]\,\},$$
where the right-hand side is a \textit{double expectation}, with the inner expectation $\E[Y_a\mid C]$ representing the mean of $Y_a$ \textit{conditional on} covariates $C$% (also referred to as \textit{given} $C$, or \textit{within levels of} $C$)
, and the outer expectation averaging the conditional mean over the covariate distribution.%
\footnote{Readers from some disciplines may be more used to writing expectations in probability-weighted average formats, so $\E_C\{\E[Y_a\mid C]\}$ may be written as $\sum_c\E[Y_a\mid C=c]\P(C=c)$ or $\sum_c\sum_yy\P(Y_a=y\mid C=c)\P(C=c)$. For complicated identification results, these formats become quite visually taxing and confusing. The expectation notation, however, remains succinct and clear on two key pieces of information: averaging what and averaging over which distribution.}
Essentially we identify $\E[Y_a]$ by identifying the conditional mean $\E[Y_a\mid C]$. To identify this conditional mean, in addition to \textit{consistency}, we need \textit{conditional independence} and \textit{positivity} assumptions. 

To understand these assumptions (formalized shortly), we need to take a close look at what is meant by covariates. For simplicity, we take covariates to be \textit{confounders}, and use the common cause definition:
% While there is not one single answer to the question what should constitute the set of covariates $C$, this paper adopts the simple (and fundamental) answer that covariates are \textit{confounders}. 
a \textit{confounder} of two variables (an exposure and an outcome) is a cause they share. As a common cause, it induces an association between the two variables, which \textit{confounds}, or confuses, their true causal relationship.
For example, education is likely a confounder of the occupation-happiness relationship, as it influences what occupation people have, and may influence happiness in ways other than through occupation. 
Confounders are only a subset of variables that can be used to remove confounding, which are called \textit{deconfounders} \citep{Pearl2014a}. The use of deconfounders that are not simply confounders is an advanced topic we leave out of this paper.

\begin{figure}[ht]
\caption{A simple DAG with variables relevant to $\E[Y_a]$}\label{fig:nomediator}
    \begin{center}
    \begin{tikzpicture}[
        box/.style={rectangle, minimum size=5mm}
    ]
        \node[box]  (C)    {$C$};
        \node[box]  (A)  [right=of C, xshift=8mm] {$A$};
        \node[box]  (Y)  [right=of A, xshift=8mm] {$Y$};
        
        \node[box]  (UA) [below=of A, yshift=+7mm] {\footnotesize $U_A$};
        \node[box]  (UY) [below=of Y, yshift=+7mm] {\footnotesize $U_Y$};
        
        \draw[->] (C) -- (A);
        \draw[->] (C) .. controls +(up:15mm) and +(up:15mm)  .. (Y);
        \draw[->] (A) -- (Y);
        
        \draw[->] (UA) -- (A);
        \draw[->] (UY) -- (Y);
    \end{tikzpicture}
    \end{center}
\end{figure}
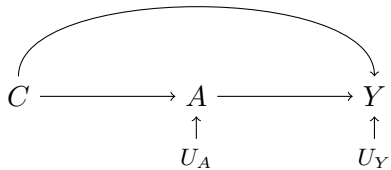

We draw a \textit{causal directed acyclic graph} (DAG) \citep{Pearl2018} in Fig. \ref{fig:nomediator} representing the relationships among the relevant variables. Because the potential outcome $Y_a$ is agnostic to mediators, we can leave mediators out of the DAG and just include: exposure $A$, outcome $Y$, common causes $C$ of these two variables, and arrows representing the causal influences among those variables.%
\footnote{The order in which a DAG is drawn is to start with variables of interest and arrows representing their causal relationships, and then add variables that are common causes of included variables.}
(In the special case where exposure is randomized, there will be no $C$, as $A$ and $Y$ do not have common causes.)
Also shown are $U_A$ and $U_Y$, causes of $A$ and $Y$ that are not shared; such unique causes are often omitted from DAGs. 
An important note: the arrow from $A$ to $Y$ in this DAG captures all the influence of $A$ on $Y$ (inclusive of influence through and not through the mediator $M$); the arrow from $C$ to $Y$ captures all the influence of $C$ on $Y$ except the part that goes through $A$.

\subsection{Conditional independence assumption}

\noindent In most applications, we are not privy to the truth, or the full truth, about confounders. What we try to do is to guess, based on prior knowledge and theory, what the important confounders are, and collect data on them. Then we resort to making the untestable assumption that the covariates $C$ we observe capture all the confounders (of the relationship between $A$ and $Y_a$). This is the gist of the \textit{conditional independence} (also known as \textit{unconfoundedness}, \textit{exchangeability} or \textit{ignorability}) assumption. 

Let us be a bit more formal here to build clarity that will help with the later potential outcome types.
For ease of reference, we label this assumption (\ref{Ia}), with I for ``independence'' and the subscript $a$ indexing potential outcome $Y_a$.
This assumption is that $Y_a$ is independent of exposure status conditional on covariates, formally
\begin{equation}
    A\independent Y_a\mid C,\tag{I$_a$}\label{Ia}
\end{equation}
where $\independent$ is the symbol for independent.
Intuitively, \textit{within each subpopulation that shares the same values on covariates $C$}, individuals are similar enough that whether an individual happens to be exposed or not does not carry any additional information about their $Y_a$. This means we can \textit{ignore} exposure status when considering $Y_a$. Put another way, within each such subpopulation, exposed and unexposed individuals are \textit{exchangeable} in the sense that they share the same distribution of variable $Y_a$.
This allows equating the subpopulation mean of $Y_a$ with the mean of $Y_a$ in the $A=a$ group in the subpopulation, formally, 
$$\E[Y_a\mid C]=\E[Y_a\mid C,A=a].$$

What conditional independence allows us to do, here and later in the paper, is what we informally call \textit{going from whole to part}, or vice versa. Here the \textit{whole} is the ($C$-value-specific) subpopulation, and the \textit{part} is the $A=a$ group in the subpopulation. The beauty of this move is that we do not need to worry about the unobserved $Y_a$ values of the individuals whose actual exposure is not $a$.

What we now want to do is to replace the potential outcome $Y_a$ on the right-hand side above with the observed $Y$. \textit{Consistency} apparently suggests doing so, since on the right-hand side, we are considering $Y_a$ only among those with $A=a$. It turns out, though, that in addition to \textit{consistency}, we also need the third assumption, \textit{positivity}.

\begin{table}[thbp]
    \centering
    \caption{Three identifying assumptions for five potential outcome means}
    \label{tab:3x5}
    \resizebox{\textwidth}{!}{%
    \begin{tabular}{@{}llllllll@{}}
         &&&
         \\[.3em]
         \multicolumn{7}{@{}l}{\textbf{CONSISTENCY ASSUMPTION}}
         \\[.3em]
         Potential outcome mean
         && 
         \begin{tabular}[b]{@{}l@{}}
             Consistency of 
             \\
             the potential outcome
         \end{tabular}
         && 
         \begin{tabular}[b]{@{}l@{}}
             Consistency of potential
             \\
             intermediate confounder
         \end{tabular}
         &&
         \begin{tabular}[b]{@{}l@{}}
             Consistency of relevant
             \\
             potential mediator(s)
         \end{tabular}
         \\ 
         \hline
         \\[-.9em]
         $\E[Y_a]$ 
         && $Y=Y_a$ if $A=a$ 
         \\
         \\[-.9em] 
         \hline
         \\[-.9em]
         $\E[Y_{am}]$ 
         && 
         $Y=Y_{am}$ if $A=a$ and $M=m$
         \\
         \\[-.9em]
         \hline
         \\[-.9em]
         \begin{tabular}[t]{@{}l@{}}
             $\E[Y_{a\M}]$, where $\M$ is a
             \\
             known distribution
         \end{tabular}
         && 
         \begin{tabular}[t]{@{}l@{}}
             For all values $m$ in dist. $\M$,\\
             $Y=Y_{am}$ if $A=a$ and $M=m$
         \end{tabular}
         && 
         \begin{tabular}[t]{@{}l@{}}
             If dist. $\M$ conditions on
             \\
             $L_a$, also require:
             \\
             $L=L_a$ if $A=a$
         \end{tabular}
         && 
         \\
         \\[-.9em] 
         \hline
         \\[-.9em]
         \begin{tabular}[t]{@{}l@{}}
             $\E[Y_{a\M}]$, where $\M$ is defined
             \\
             based on dist. of $M_{a^*}$
         \end{tabular}
         && 
         Same as above
         && 
         \begin{tabular}[t]{@{}l@{}}
             Same as above, and 
             \\[.5em]
             If definition of $\M$ relies
             \\
             on info. about dist. of
             \\
             $M_{a^*}$ given $L_{a^*}$, also require:
             \\
             $L=L_{a^*}$ if $A=a^*$
         \end{tabular}
         && $M=M_{a^*}$ if $A=a^*$
         \\
         \\[-.9em] 
         \hline
         \\[-.9em]
         $\E[Y_{aM_{a'}}]$, where $a\neq a'$ 
         && 
         \begin{tabular}[t]{@{}l@{}}
             For all $m$ in dist. of $M_{a'}\mid C$, 
             \\
             $Y=Y_{am}$ if $A=a$ and $M=m$;
             \\
             $Y_{aM_{a'}}=Y_{am}$ if $M_{a'}=m$
         \end{tabular}
         &&
         && $M=M_{a'}$ if $A=a'$
         \\
         \\[-.9em]
         \hline
         \\[-.5em]
         \multicolumn{7}{@{}l}{\textbf{CONDITIONAL INDEPENDENCE ASSUMPTION}} 
         \\[.3em]
         Potential outcome mean
         && 
         \begin{tabular}[b]{@{}l@{}}
             Exposure-outcome 
             \\
             conditional independence
         \end{tabular}
         && 
         \begin{tabular}[b]{@{}l@{}}
             Mediator-outcome
             \\
             conditional independence
         \end{tabular}
         &&
         \begin{tabular}[b]{@{}l@{}}
             Exposure-mediator
             \\
             conditional independence
         \end{tabular}
         \\ 
         \hline
         \\[-.9em]
         $\E[Y_a]$ 
         && $A\independent Y_a\mid C$
         \\
         \\[-.9em] 
         \hline
         \\[-.9em]
         $\E[Y_{am}]$ 
         && $A\independent Y_{am}\mid C$
         && $M\independent Y_{am}\mid C,L,A=a$
         \\
         \\[-.9em]
         \hline
         \\[-.9em]
         \begin{tabular}[t]{@{}l@{}}
             $\E[Y_{a\M}]$, where $\M$ is a\\
             known distribution
         \end{tabular}
         &&
         \begin{tabular}[t]{@{}l@{}}
             For all values $m$ in dist. $\M$,
             \\
             $A\independent Y_{am}\mid C$
            %  \\[.5em]
            %  If dist. $\M$ conditions on $L_a$,
            %  \\
            %  replace with:
            %  \\
            %  $A\independent(L_a,Y_{am})\mid C$
         \end{tabular}
         &&
         \begin{tabular}[t]{@{}l@{}}
             For all values $m$ in dist. $\M$,
             \\
             $M\independent Y_{am}\mid C,L,A=a$
         \end{tabular}
         \\
         \\[-.9em]
         \hline
         \\[-.9em]
         \begin{tabular}[t]{@{}l@{}}
             $\E[Y_{a\M}]$, where $\M$ is defined 
             \\
             based on dist. of $M_{a^*}$
         \end{tabular}
         && 
         Same as above
         && 
         Same as above
         && 
         \begin{tabular}[t]{@{}l@{}}
             $A\independent M_{a^*}\mid C$ 
            %  \\[.5em]
            %  If rely on dist. of $M_{a^*}$
            %  \\
            %  given $L_{a^*}$, replace with:
            %  \\
            %  $A\independent (L_{a^*},M_{a^*})\mid C$
         \end{tabular}
         \\
         \\[-.9em]
         \hline
         \\[-.9em]
         $\E[Y_{aM_{a'}}]$, where $a\neq a'$ 
         && 
         \begin{tabular}[t]{@{}l@{}}
             For all $m$ in dist. of $M_{a'}\mid C$,\\
             $A\independent Y_{am}\mid C$
         \end{tabular}
         && 
         \begin{tabular}[t]{@{}l@{}}
             For all $m$ in dist. of $M_{a'}\mid C$,
             \\[.3em]
             $M\independent Y_{am}\mid C,A=a$ and
             \\
             $M_{a'}\independent Y_{am}\mid C$
            %  \\[.3em]
            %  or more simply,
            %  \\[.3em]
            %  $M_1,M_0\independent Y_{am}\mid C,A$
         \end{tabular}
         && $A\independent M_{a'}\mid C$
         \\
         \\[-.9em]
         \hline
         \\[-.5em]
         \multicolumn{7}{@{}l}{\textbf{POSITIVITY ASSUMPTION}}
         \\[.3em]
         Potential outcome mean
         && Positivity of exposure conditions
         && Positivity of mediator values
         && 
        %  \begin{tabular}[b]{@{}l@{}}
        %      Positivity of
        %      \\
        %      intermediate confounders
        %  \end{tabular}
         \\ 
         \hline
         \\[-.9em]
         $\E[Y_a]$ 
         && $\P(A=a\mid C)>0$ 
         \\
         \\[-.9em] 
         \hline
         \\[-.9em]
         $\E[Y_{am}]$ 
         && 
         Same as above
         && $\P(M=m\mid C,L,A=a)>0$
         \\
         \\[-.9em]
         \hline
         \\[-.9em]
         \begin{tabular}[t]{@{}l@{}}
             $\E[Y_{a\M}]$, where $\M$ is
             \\
             a known distribution
         \end{tabular}
         &&
         Same as above
         &&
         \begin{tabular}[t]{@{}l@{}}
             For all values $m$ in dist. $\M$,\\
             $\P(M=m\mid C,L,A=a)>0$
         \end{tabular}
         \\
         \\[-.9em] 
         \hline
         \\[-.9em]
         \begin{tabular}[t]{@{}l@{}}
             $\E[Y_{a\M}]$, where $\M$ is defined
             \\
             based on dist. of $M_{a^*}$
         \end{tabular}
         &&
         \begin{tabular}[t]{@{}l@{}}
             Same as above, and
             \\[.5em]
             if $a^*\neq a$, also require
             \\
             $\P(A=a^*\mid C)>0$
             \\[.7em]
             if $a^*\neq a$ and dist. $\M$ conditions
             \\
             on $(C,L_a)$ and is defined to be
             \\
             the same as dist. of $M_{a^*}$ given
             \\
             $(C,L_{a^*})$, also require
             \\
             $\P(A=a^*\mid C,L)>0$
         \end{tabular}
         && 
         Same as above
         &&
        %  \begin{tabular}[t]{@{}l@{}}
        %      If dist. $\M$ conditions on
        %      \\
        %      $(C,L_a)$ and is defined to
        %      \\
        %      be the same as dist. of
        %      \\
        %      $M_{a^*}$ given $(C,L_{a^*})$, 
        %      \\
        %      also require:
        %      \\
        %      $\P(L=l\mid C,A=a^*)>0$
        %      \\
        %      for all $l$ in dist. of $L_a\mid C$
        %      \\
        %      \textcolor{red}{REPLACE W/ SIMPLER}
        %      \\
        %      $\P(A=A^*\mid C,L)>0$
        %      \\
        %      \textcolor{red}{AND ADD PROOF}
        %  \end{tabular}
         \\
         \\[-.9em] 
         \hline
         \\[-.9em]
         $\E[Y_{aM_{a'}}]$, where $a\neq a'$ 
         && $0<\P(A=a\mid C)<1$
         &&
         \begin{tabular}[t]{@{}l@{}}
             For all $m$ in dist. of $M_{a'}\mid C$,\\
             $\P(M=m\mid C,A=a)>0$
         \end{tabular}
         \\
         \\[-.9em]
         \hline
         \\[-.9em]
         \multicolumn{7}{@{}l}{Note: dist. = distribution.}
    \end{tabular}%
    }
\end{table}

\subsection{Positivity assumption}

\noindent A replacement of the $\E[Y_a\mid C,A=a]$ above with $\E[Y\mid C,A=a]$ is only legitimate if the latter is well defined for all values of $C$. This requires the assumption that there is a positive chance of $A=a$ for all $C$ values, formally, $\P(A=a\mid C)>0$, where $\P(\cdot)$ is the notation for probability or probability density. 
Combined with \textit{consistency}, this means that for all $C$ values there is a positive chance of observing $Y_a$. If this is not the case for certain $C$ values, then the mean of $Y_a$ given those values is unidentified, and thus $\E[Y_a]$ is unidentified.

Unlike the other two assumptions, \textit{positivity} is testable, in the sense that given data that have been collected, one could check whether there are parts of the observed covariate distribution where there are no individuals with exposure condition $a$.

\smallskip

For ease of reference, these three assumptions for all of the five potential outcome types are collected in Table \ref{tab:3x5}.

\subsection{Identification result}

\noindent The three assumptions combined help identify the conditional mean of $Y_a$,
$$\E[\textcolor{purple}{Y_a}\mid C]\overset{\scriptsize\begin{tabular}{c}conditional\\independence\end{tabular}}{=}\E[\textcolor{purple}{Y_a}\mid C, A=a]\overset{\scriptsize{\begin{tabular}{c}consistency and\\positivity\end{tabular}}}{=}\E[Y\mid C,A=a].$$
And then using double expectation, we identify the population mean:
\begin{equation}
    \E[\textcolor{purple}{Y_a}]=\E_C\{\,\E[Y\mid C,A=a]\,\}.\tag{R$_a$}\label{Ra}
\end{equation}
Here the inner expectation is the mean \textit{observed} outcome given $C$ among those in the $A=a$ condition, and the outer expectation averages this over the distribution of $C$.
(In identification result labels, R stands for ``result''.)

% \medskip

% In place of a summary, to help connect with later sets of assumptions, we note the themes to which the three assumptions above belong: \textit{consistency of the potential outcome}, \textit{exposure-outcome conditional independence}, and \textit{positivity of exposure condition}.

\subsection{Application 1: the total effect}

\noindent Identifying the average total effect, $\mathrm{TE}=\E[Y_1]-\E[Y_0]$, involves identifying both potential outcome means. The assumptions required, collected in Table \ref{tab:TE}, include (i) consistency of both potential outcomes, (ii) conditional independence of exposure with both potential outcomes; and (iii) positivity of both exposure conditions. (i) and (iii) are often written concisely as $Y=AY_1+(1-A)Y_0$ and $0<\P(A=1\mid C)<1$, respectively. Under these assumptions, the identification result is 
$\text{TE}=\E_C\big\{\E[Y\mid C,A=1]\big\}-\E_C\big\{\E[Y\mid C,A=0]\big\}$.

\begin{table}[h]
    \centering
    \caption{Identifying assumptions for TE}
    \label{tab:TE}
    \scriptsize
    \begin{tabular}{llcc}
        && \multicolumn{2}{l}{Relevant to}
        \\
        & Assumptions 
        & $\E[Y_1]$ & $\E[Y_0]$  
        \\\hline
        Consistency
        & $Y=Y_1$ if $A=1$
        & \checkmark
        \\
        & $Y=Y_0$ if $A=0$
        && \checkmark
        \\\hline
        Conditional independence
        & $A\independent Y_1\mid C$
        & \checkmark
        \\
        & $A\independent Y_0\mid C$
        && \checkmark
        \\\hline
        Positivity
        & $\P(A=1\mid C)>0$
        & \checkmark
        \\
        & $\P(A=0\mid C)>0$
        && \checkmark
        \\\hline
    \end{tabular}
\end{table}

In the example, we consider all effects as average effects over the population of people with a bipolar diagnosis and a chronic medical problem who are in the local branch of the health care system. The total effect is difference between the potential outcome means under intervention and under usual care, where the means are taken over this population.

The consistency assumption simply says that the observed quality of life in a patient in the intervention group is the same as their potential quality of life under intervention ($Y_1$), and the observed quality of life in a patient in the usual care group is the same as their potential quality of life under usual care ($Y_0$). 

We assemble the set $C$ of likely confounders (i.e., common causes of intervention participation and quality of life): age, sex, education, occupation, income, psychiatric and medical diagnoses, baseline psychiatric and medical symptoms, baseline quality of life, and baseline measures of self management of symptoms and effective medical care use.
The conditional independence assumption says that individuals that share the same values on these $C$ variables also share the same $Y_1$ distribution and the same $Y_0$ distribution, regardless of whether they actually receive the intervention or usual care. If an important confounder, say baseline quality of life, is not included in $C$, this assumption is violated. 

The positivity assumption means that for individuals with any given realization of $C$, there is a positive chance of receiving the intervention and a positive chance of receiving usual care. If the sample includes individuals with some specific covariate value none of whom participated in the intervention (e.g., patients who could not attend group sessions due to physical mobility challenges), positivity is violated.  

A couple of comments: First, in most practical settings, the distinction between the double conditional independence assumption for TE (which involves both $Y_1$ and $Y_0$) and the single version specific to one potential outcome $Y_a$ may not matter, for we would arrive at roughly the same set of covariates. This is fortunate, as substantive considerations tend to be imprecise -- asking what are the common causes of $A$ and $Y$, instead of $A$ and $Y_a$ for a specific value $a$. Note though that when we need to identify the mean of $Y_0$ or of $Y_1$ but not both (e.g., the effect of a modified intervention relative to usual care involves $Y_0$ but not $Y_1$), only the single version relevant to the specific potential outcome is required. 
Second, the reason why the double positivity assumption here is called \textit{covariate overlap} or \textit{common support} is that positivity of both exposure conditions implies that the support of the covariate distribution (i.e., range of covariate values) is shared between the exposed and unexposed.

\section{Several types of confounders in the mediation setting}

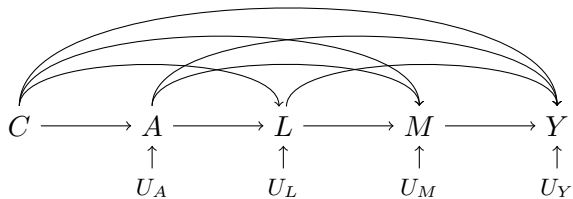
\begin{figure}[h]
\caption{A shorthand DAG: $C$ is a collection of variables, each of which has at least two of the four depicted arrows emitted from $C$}\label{fig:main}
    \begin{center}
    \begin{tikzpicture}[
        box/.style={rectangle, minimum size=5mm}
    ]
        \node[box]  (C)    {$C$};
        \node[box]  (A)  [right=of C, xshift=2mm] {$A$};
        \node[box]  (L)  [right=of A, xshift=2mm] {$L$};
        \node[box]  (Lleft)  [right=of A, xshift=1.5mm] {};
        \node[box]  (Lright)  [right=of A, xshift=2.5mm] {};
        \node[box]  (M)  [right=of L, xshift=2mm] {$M$};
        \node[box]  (Y)  [right=of M, xshift=2mm] {$Y$};
        
        \node[box]  (UA) [below=of A, yshift=+7mm] {\footnotesize $U_A$};
        \node[box]  (UL) [below=of L, yshift=+7mm] {\footnotesize $U_L$};
        \node[box]  (UM) [below=of M, yshift=+7mm] {\footnotesize $U_M$};
        \node[box]  (UY) [below=of Y, yshift=+7mm] {\footnotesize $U_Y$};
        
        \draw[->] (C) -- (A);
        \draw[->] (C) .. controls +(up:10mm) and +(up:10mm)  .. (Lleft);
        \draw[->] (C) .. controls +(up:15mm) and +(up:15mm)  .. (M);
        \draw[->] (C) .. controls +(up:20mm) and +(up:20mm)  .. (Y);
        \draw[->] (A) -- (L);
        \draw[->] (A) .. controls +(up:10mm) and +(up:10mm)  .. (M);
        \draw[->] (A) .. controls +(up:15mm) and +(up:15mm)  .. (Y);
        \draw[->] (L) -- (M);
        \draw[->] (Lright) .. controls +(up:10mm) and +(up:10mm)  .. (Y);
        \draw[->] (M) -- (Y);
        
        \draw[->] (UA) -- (A);
        \draw[->] (UL) -- (L);
        \draw[->] (UM) -- (M);
        \draw[->] (UY) -- (Y);
    \end{tikzpicture}
    \end{center}
\end{figure}

\noindent Unlike $Y_a$, the other four potential outcome types correspond to conditions where both the exposure and the mediator are manipulated. Since the mediator is manipulated, the relevant DAG is expanded from the one shown in Fig. \ref{fig:nomediator}. It includes exposure $A$, outcome $Y$, mediator of interest $M$, and covariates that are common causes of any two of these variables. 
Intuitively, we can think of the covariates as consisting of confounders of the exposure-mediator, exposure-outcome and mediator-outcome relationships. (And in the special case where exposure is randomized, there are only mediator-outcome confounders.) 
For any application, the actual DAG (the one that informs the analysis) may get rather complicated, with multiple confounders and complex causal relationships. For the sake of explaining the identifying assumptions, the causal mediation literature commonly uses a shorthand DAG of the form in Fig. \ref{fig:main} \citep[e.g.,][]{VanderWeele2014a}.

Here covariates are represented by $C$ and $L$, which differentiate whether they are influenced by exposure -- $L$ is but $C$ is not. Relating this to the relationships to be confounded, exposure-mediator and exposure-outcome confounders all belong in $C$ -- they influence exposure rather than the other way around. Mediator-outcome confounders are split between $C$ and $L$: those not influenced by exposure (regardless of whether they influence exposure or not) are in $C$, while those influenced by exposure are in $L$. We call this DAG shorthand because $C$ is a collection of different (although overlapping) types of variables. Precisely, of the four arrows depicted as emitting from the node $C$ in the DAG to the $A,L,M,Y$, each variable in $C$ needs to have a minimum of only two.%
\footnote{Each exposure-mediator confounder has an arrow into $A$ and at least an arrow into $L$ or $M$. Each exposure-outcome confounder has an arrow into $A$ and at least an arrow into $L$, $M$ or $Y$. Each mediator-outcome confounder in $C$ has at least two of the three arrows into $L$, $M$ and $Y$.}

Note that the current set $C$ is larger than the set $C$ in Fig. \ref{fig:nomediator}. The set in Fig. \ref{fig:nomediator} consists of exposure-outcome confounders (which include exposure-mediator confounders%
\footnote{Exposure-mediator confounders form a subset of exposure-outcome confounders, because a cause of the mediator is also a cause of the outcome (via the mediator).}%
),
but does not include mediator-outcome confounders that do not influence exposure. While an abuse of notation, the reuse of the label $C$ here is not problematic because the current set $C$ also works for the identifying assumptions of $\E[Y_a]$. In fact, there are a couple of other places where either the current set $C$ or a subset of it could be used in an assumption. To keep presentation simple, we simply use $C$ when stating the assumptions, but also note subsets that could replace $C$ where appropriate.

These $C$ variables are commonly referred to as \textit{pre-exposure confounders} or \textit{pre-exposure covariates}. This terminology is somewhat imprecise, as mediator-outcome confounders in $C$ do not necessarily precede exposure either in time or in the causal structure; the key point is that they are not influenced by exposure. $L$ variables are often called \textit{post-exposure confounders} or \textit{intermediate confounders}. 
The latter label signals that they are intermediate variables, i.e., also mediators of the effect of $A$ on $Y$, albeit not the mediator of interest ($M$). Another way to think of $L$ variables is that they are variables on the causal pathway from $A$ to $M$ that happen to also influence $Y$ in ways that are not through $M$.

% In the literature, $C$ in the current DAG is usually referred to, somewhat imprecisely, as \textit{pre-exposure confounders} or \textit{pre-exposure covariates}. This is perhaps due to a general concern that variables that follow exposure in time are potentially influenced by exposure, and if so, would belong in $L$. Following convention, we adopt this convenient \textit{pre-exposure} label for $C$, but note that for mediator-outcome confounders, the key distinction that puts them in $C$ vs. $L$ is whether they are influenced by exposure or not, not whether they precede or follow exposure in time.
% As for $L$ variables, they are commonly called \textit{post-exposure confounders} or \textit{intermediate confounders}. The latter label signals that these are intermediate variables, that is, they are themselves mediators of the effect of $A$ on $Y$, albeit not the mediator of interest ($M$). Another helpful way to think of $L$ variables is that they are variables on the causal pathway from $A$ to $M$ that happen to also influence $Y$ in ways that are not through $M$.

In the illustrative example with the intervention for bipolar patients, depending on the specific research question, one may take symptom management, or service use, or the combination of both, as the mediator of interest $M$. Whichever the choice, the covariates need to be expanded to cover the different types of confounders. In principle, exposure-mediator confounders should already be part of the exposure-outcome confounders selected when considering the total effect, but it is helpful to double-check. Also, mediator-outcome confounders may need to be added (as $C$ and $L$ variables); and this should be thought through for the specific mediator being considered.

If symptom management is taken as $M$, baseline general health-related self-efficacy may be an important common cause of symptom management and quality of life. We thus add this variable to the pre-exposure covariate set $C$. Since symptom management is measured early on (at completion of the self-management intervention component), we are confident that there are no important post-exposure confounders.

If service use is taken as $M$, then symptom management is likely an intermediate confounder (an $L$ variable). In addition, we add to the set $C$ a variable indicating whether the patient had an annual checkup in the previous year, which is believed to reflect a baseline tendency to use medical services for self care, a likely mediator-outcome confounder. A variable indicating the patient's specific health insurance plan is also added to the set $C$.

\section{Identification of $\E[Y_{am}]$}

\subsection{Consistency assumption}

\noindent For $\E[Y_{am}]$, this assumption is simply that $Y=Y_{am}$ if $A=a$ and $M=m$ (that is, in individuals with actual exposure $a$ and actual mediator value $m$, their observed $Y$ reveals their potential outcome $Y_{am}$), for $a$ and $m$ being the specific exposure and mediator values that define the potential outcome.

\subsection{Conditional independence assumption}

\noindent Recall that $\E[Y_a]$ identification requires \textit{exposure-outcome} conditional independence. Identification of $\E[Y_{am}]$, which corresponds to a condition where not only the exposure but also the mediator is manipulated, requires an assumption of both \textit{exposure-outcome} and \textit{mediator-outcome} conditional independence, where \textit{outcome} refers to $Y_{am}$. Specifically,

\begin{align}
    A&\independent Y_{am}\mid C,\tag{I$_{am}$-\textsc{ay}}\label{Iam-ay}
    \\
    M&\independent Y_{am}\mid C,L,A=a.\tag{I$_{am}$-\textsc{my}}\label{Iam-my}
\end{align}

\medskip

Here the exposure-outcome component (\ref{Iam-ay}) says that within levels of $C$, individuals are similar enough that their actual exposure condition provides no information about $Y_{am}$. The mediator-outcome component (\ref{Iam-my}) says that among those with exposure $A=a$, within levels of the combination of covariates $\{C,L\}$, individuals are similar enough that the actual mediator value $M$ does not provide any additional information about $Y_{am}$. (In other words, with appropriate conditioning, exposure is \textit{as good as randomized} and so is mediator value.) The exposure-outcome component is similar to assumption (\ref{Ia}), and like (\ref{Ia}), it is satisfied by design if exposure is randomized. Since the mediator is not randomized, the mediator-outcome component is always an untestable assumption.

Relating to other terminology, this assumption (essentially \textit{unconfoundedness of $Y_{am}$}) includes \textit{no unobserved exposure-outcome and mediator-outcome confounding}, where $C$ captures all $A$-$Y_{am}$ confounders, and $C$ and $L$ capture all $M$-$Y_{am}$ confounders for those with $A=a$. The two components of this assumption could also be referred to as \textit{ignorability of exposure assignment}, and \textit{ignorability of mediator value assignment}, for $Y_{am}$. In \textit{exchangeability} terms, the distribution of $Y_{am}$ is exchangeable between exposure conditions conditional on $C$, and is exchangeable across mediator values within the $A=a$ group conditional on $C,L$.

\subsection{Positivity assumption}

\noindent Because $Y_{am}$ corresponds to a condition in which both exposure and mediator are manipulated, the positivity assumption includes both \textit{positivity of exposure condition $a$} and \textit{positivity of mediator value $m$}. The latter is formally $\P(M=m\mid C,L,A=a)>0$. That is, among those with $A=a$, the chance of $M=m$ is positive for any combination of values that covariates $\{C,L\}$ may take.

\medskip

A side note about $L$: In both the mediator conditional independence and mediator positivity assumptions above, $L$ is part of the conditioning set. We will see that for this and the next two potential outcome types (and all the interventional effects that involve them), identification is possible with intermediate confounders $L$, but for the fifth potential outcome type (and the natural (in)direct effects that involve it) the presence of $L$ generally results in nonidentifiability.

\subsection{Identification}

\noindent Under \textit{consistency}, $Y_{am}$ values are observed in individuals with $A=a$ and $M=m$. To see how the other assumptions then bridge to the population mean of $Y_{am}$, it is easier to first consider the special case with no intermediate confounders. In this special case, (\ref{Iam-my}) reduces to $M\independent Y_{am}\mid C,A=a$, and the argument involves \textit{going from whole to part} twice. Again, consider a subpopulation of individuals that share the same values of $C$. In this subpopulation, the overall mean of $Y_{am}$ is equal to the mean of $Y_{am}$ in those with $A=a$ (because exposure status is ignorable for $Y_{am}$ under (\ref{Iam-ay})), which in turn is equal to the mean of $Y_{am}$ among those with $A=a$ and $M=m$ (because mediator value is ignorable for $Y_{am}$ among those with $A=a$ under (\ref{Iam-my})). Formally,
\begin{align*}
    \E[\textcolor{purple}{Y_{am}}\mid C]
    \overset{\scriptsize{\begin{tabular}{c}conditional\\independence\\(\ref{Iam-ay})\end{tabular}}}{=}
    &\E[\textcolor{purple}{Y_{am}}\mid C,A=a]
    \overset{\scriptsize{\begin{tabular}{c}conditional\\independence\\(\ref{Iam-my})\end{tabular}}}{=}\E[\textcolor{purple}{Y_{am}}\mid C,A=a,M=m].
\end{align*}
Similar to previous reasoning, under \textit{consistency} and \textit{positivity}, the right-hand side is replaced with $\E[Y\mid C,A=a,M=m]$, thus identifying the conditional mean $\E[Y_{am}\mid C]$. Then using the double expectation trick, we obtain
$\E[\textcolor{purple}{Y_{am}}]=\E_C\{\,\E[Y\mid C,A=a,M=m]\,\}$,
where the inner expectation is the mean \textit{observed} outcome among those with $A=a,M=m$ within levels of $C$, and the outer expectation averages over the distribution of $C$. This is very much in the same spirit as (\ref{Ra}), the identification result of $\E[Y_a]$.

In the general case with intermediate confounders ($L$), the bridging involves additional steps similar in nature to those above. We leave the details to the Appendix and just consider the result here. As the outcome depends on $L$ in addition to exposure, mediator, and $C$, the inner expectation now conditions on both $C$ and $L$, becoming $\E[Y\mid C,L,A=a,M=m]$, the mean observed outcome for those with $A=a,M=m$ within levels of the combination of $\{C,L\}$. And instead of the double expectation, we have a triple expectation that averages over $L$ before averaging over $C$, where the distribution of $L$ that is averaged over is that of those with $A=a$, within levels of $C$.
Our general identification result is
\begin{align}
    \E[\textcolor{purple}{Y_{am}}]=\E_C(\,\E_{L\mid C,A=a}\{\,\E[Y\mid C,L,A=a,M=m]\,\}\,).\tag{$\mathrm{R}_{am}$}\label{Ram}
\end{align}

\medskip

There are two observations that are technical (and non-essential) but may provide additional insight. First, since $\E[Y_{am}]$ concerns only one mediator value, $m$, (\ref{Iam-my}) may be simplified by replacing $M$ with a dichotomized version of this variable indicating whether it is equal to $m$ or not. We can thus think about $Y_{am}$ as the potential outcome in a condition that intervenes on two binary variables. In most applications this likely does not make a difference as to which variables are included in $C$ and $L$. However, this clarity might help make the assumption more meaningful%
\footnote{For some reason, it seems that the idea of ignorability of exposure assignment is quite intuitive, but the idea of ignorability of mediator value assignment is illusive. This puts them both on equal footing.}
as it provides a parallelism: (\ref{Iam-ay}) says within levels of $C$, individuals are similar enough that whether they receive $A=a$ or not does not carry any information about $Y_{am}$; and the simplified (\ref{Iam-my}) says that in the $A=a$ condition, within levels of the combination of $\{C,L\}$, individuals are similar enough that whether they receive $M=m$ or not does not tell us anything about $Y_{am}$.

Second, $C$ here may be replaced by a subset of $C$ consisting of variables that directly influence $L$ and/or $Y$ (labeled $C^{LY}$), leaving out variables with no arrows to $L$ and $Y$ (labeled $C^{\bcancel{LY}}$).
The reason is simple: when targeting $Y_{am}$ we need to deal with confounders of the relationship between the combination $\{A,M\}$ and $Y_{am}$, and variables in $C^{\bcancel{LY}}$ influence the former but do not influence the latter (other than through the former) so they do not confound this relationship. Intuitively, variables that are included in $C$ \textit{only because} they are exposure-mediator confounders may be ignored for the purpose of identifying $\E[Y_{am}]$.% 
\footnote{If we consider the two components of (I$_{am}$) and the three elements of (\ref{Ram}) separately, the answer is: (i) $C$ in (\ref{Iam-ay}) and in the outer expectation in (\ref{Ram}) can be replaced with $C^{LY}$; (ii) $C$ in (\ref{Iam-my}) and in the inner expectation in (\ref{Ram}) can be replaced with a subset of $C^{LY}$ that additionally leaves out variables with an arrow to $L$ that without this arrow would not have been included in $C$; and (iii) $C$ in the middle expectation in (\ref{Ram}) can be replaced with another subset of $C^{LY}$ that contains only $A$-$L$ confounders.}

\subsection{Application 2: a controlled direct effect}

\noindent Now suppose that our local branch of the health care system receives communication from headquarters that leadership is considering a system-wide revamping of standard operating procedures which would incorporate substantial support for the care of medical problems in people with psychiatric disorders, through the use of a range of case management, provider education, integrated patient records and enhanced linkage network solutions. The communication also says that the expected result of this practice change is to obtain the ``maximal effectiveness in use of medical services for care of chronic medical conditions''.

Such a system change would have many implications which branch management has to consider. One of the first questions asked is, if left as is, what would be the effect of the intervention for bipolar patients in the new context, where (based on the expected result of the system change) the service use variable takes the highest value (5 on a 0-to-5 scale) for all bipolar patients.
Treating this variable as the mediator ($M$), the question points to the controlled direct effect $\text{CDE}(5)=\E[Y_{1,5}]-\E[Y_{0,5}]$ for mediator control level 5. (There is, however, doubt about whether this high level is realistic.)

CDE(5) identification requires identifying the means of $Y_{1,5}$ and $Y_{0,5}$, with assumptions collected in Table \ref{tab:CDE}. Note that within the conditional indepence assumption, the mediator-outcome component is exposure congruent. It is among patients in the \textit{intervention} program ($A=1$) that we assume service use ($M$) is ignorable for the potential outcome under \textit{intervention}-and-highest-service-use ($Y_{1,5}$) given baseline covariates ($C$) and symptom management ($L$).
% it is assumed that conditional on baseline covariates and symptom management ($C,L$), the service use score ($M$) is ignorable for the potential outcome under \textit{intervention} and under level 5 service use ($Y_{1,5}$).
Similarly, conditional independence of $M$ with $Y_{0,5}$ is required among patients in usual care ($A=0$) only.

\begin{table}[h]
    \centering
    \caption{Identifying assumptions for CDE(5)}
    \label{tab:CDE}
    \scriptsize
    \begin{tabular}{lllcc}
        &&& \multicolumn{2}{l}{Relevant to}
        \\
        && Assumptions 
        & $\E[Y_{1,5}]$ & $\E[Y_{0,5}]$
        \\\hline
        \multicolumn{2}{l}{Consistency}
        & $Y=Y_{1,5}$ if $A=1,M=5$
        & \checkmark
        \\
        && $Y=Y_{0,5}$ if $A=0,M=5$
        && \checkmark
        \\\hline
        \multicolumn{2}{l}{Conditional independence:}
        \\
        & exposure-outcome
        & $A\independent Y_{1,5}\mid C$
        & \checkmark
        \\
        && $A\independent Y_{0,5}\mid C$
        && \checkmark
        \\
        & mediator-outcome
        & $M\independent Y_{1,5}\mid C,L,A=1$
        & \checkmark
        \\
        && $M\independent Y_{0,5}\mid C,L,A=0$
        && \checkmark
        \\\hline
        \multicolumn{2}{l}{Positivity:}
        \\
        & exposure condition 
        & $\P(A=1\mid C)>0$
        & \checkmark
        \\
        && $\P(A=0\mid C)>0$
        && \checkmark
        \\
        & mediator value
        & $\P(M=5\mid C,L,A=1)>0$
        & \checkmark
        \\
        && $\P(M=5\mid C,L,A=0)>0$
        && \checkmark
        \\\hline
    \end{tabular}
\end{table}

% Assumptions (i), (ii-a) and (iii-a) are similar for those required for the TE, except they concern a different pair of potential outcomes. Assumption mediator-outcome conditional independence assumption (ii-b) means that within the intervention group, once we have conditioned on baseline covariates ($C$) and symptom management score ($L$), the actual effective service use score ($M$) does not tell us anything about $Y_{1,5}$ (potential quality of life under intervention and under the highest level of effective service use). Similarly, within the usual care group, conditional on $C$ and $L$, the actual $M$ value does not tell us anything about $Y_{0,5}$.

The positivity of mediator value assumption, written concisely $\P(M=5\mid C,A,L)>0$, means that in either exposure condition (intervention or usual care) patients with any realized values of $\{C,L\}$ had a positive chance of scoring 5 on service use. This assumption is violated if there is any subpopulation defined by $C,L,A$ values whose range of this variable does not include level 5.

These assumptions in Table \ref{tab:CDE} are weaker than assumptions often stated for controlled direct effects in the literature \citep[e.g.,][]{VanderWeele2009} in two ways. First, our assumptions involve only the specific mediator control level ($m=5$), while assumptions in the literature cover all possible mediator values. Such blanket assumptions are less likely to hold, and are only needed to identify the collection of controlled direct effects corresponding to every mediator level. Second, the mediator-outcome conditional independence assumption in the literature is not exposure specific, e.g., $M\independent Y_{am}\mid C,L,A$ (for $a=1,0$). For $Y_{1m}$ for example, this statement means both $M\independent Y_{1m}\mid C,L,A=0$ and $M\independent Y_{1m}\mid C,L,A=1$. We only require the latter.

Under the assumptions in Table \ref{tab:CDE}, $\text{CDE}(5)$ is identified by
$\E_C\big(\E_{L\mid C,A=1}\big\{\E[Y\mid C,L,A=1,M=5]\big\}\big)-\E_C\big(\E_{L\mid C,A=0}\big\{\E[Y\mid C,L,A=0,M=5]\big\}\big)$, a straightforward application of (\ref{Ram}).

\section{Identification of $\E[Y_{a\M}]$ where $\M$ is a known distribution} 

\noindent The assumptions that identify $\E[Y_{a\M}]$ for a known distribution $\M$ build on those that identify $\E[Y_{am}]$. The key extension is that the assumptions are now required to hold for all values $m$ in the support of the distribution $\M$.
Note that if the interventional mediator distribution $\M$ is defined conditional on covariates, the range of $m$ values depends on the covariates. For example, if $\M$ represents an intervention that differentiates by sex, then the range of $m$ for which the assumptions must hold may differ between men and women.

\subsection{Consistency assumption}

\noindent \textit{Consistency of the potential outcome} now means $Y=Y_{am}$ if $A=a$ and $M=m$ for all relevant values $m$ in the support of the distribution $\M$.
In addition, if the distribution $\M$ is defined conditional on post-exposure covariates $L_a$, then \textit{consistency of $L_a$} is also required, that is, $L=L_a$ if $A=a$.

\subsection{Conditional independence assumption}

\noindent This assumption%
\footnote{If the distribution $\M$ is defined conditional on $L_a$, strictly speaking, the first component of the assumption is $A\independent (L_a,Y_{am})\mid C$. Conditional independence between $A$ and $L_a$ is needed because, since $L_a$ is observed only in those with $A=a$, identification of $\E[Y_{a\M}]$ (where $\M$ conditions on $L_a$) requires borrowing information about $L_a$ across exposure conditions. Fortunately this distinction does not make any practical difference, because the $C$ variables we are considering are confounders. A set of confounders $C$  that satisfies $A\independent Y_{am}\mid C$ also satisfies $A\independent L_a\mid C$. This follows from the fact that $L_a$ is a cause of $Y_{am}$, which means the confounders of the $A$-$Y_{am}$ relationship (a subset of $C$) contain the confounders of the $A$-$L_a$ relationship.\label{fn10}}
is that for all the relevant values $m$ in the support of the distribution $\M$,
\begin{align}
    A&\independent Y_{am}\mid C,\tag{I$_{a\M}$-\textsc{ay}}\label{IaM-ay}
    \\
    M&\independent Y_{am}\mid C,L,A=a.\tag{I$_{a\M}$-\textsc{my}}\label{IaM-my}
\end{align}
This assumption calls for conditional independence of exposure and mediator with not just one potential outcome corresponding to a single mediator value, but a collection of potential outcomes corresponding to the range of mediator values from the distribution $\M$.
Like the previous case, $C$ here may be replaced by the subset $C^{LY}$.

\subsection{Positivity assumption}

\noindent This assumption includes both \textit{positivity of exposure condition $a$} and \textit{positivity of relevant mediator values}. The latter is $\P(M=m\mid C,L,A=a)$ for all values $m$ in the support of $\M$. This means among those with exposure $a$, within each subpopulation that shares the same $\{C,L\}$ values, the actual range of mediator values has to cover the range of values prescribed by the interventional distribution $\M$. If not, this assumption is violated.

\subsection{Identification result}

\noindent The identification result is an extension of the triple expectation (\ref{Ram}) to a quadruple expectation that involves averaging over the distribution $\M$ in addition to averaging over $L$ and $C$.
\begin{align}
    \E[\textcolor{purple}{Y_{a\M}}]=\E_C[\,\E_{L\mid C,A=a}(\,\E_{\M}\{\,\E[Y\mid C,L,M,A=a]\,\}\,)\,].\tag{R$_{a\M}$}\label{RaM}
\end{align}

\subsection{Application 3: a generalized direct effect}\label{app3}

\noindent Additional communication from headquarters later clarified that the earlier statement about ``maximal effectiveness in use of medical services'' meant matching the effective service use level of otherwise similar patients who are without psychiatric disorders, not the highest possible level 5. This means that we are drawing information from the observed effective service use distribution in non-psychiatric patients, conditional on key covariates (age, sex, education, occupation, income, health insurance plan, medical diagnoses, previous year annual checkup).
Instead of the controlled direct effect CDE(5), we are now considering the generalized direct effect $\text{GDE}(\M)=\E[Y_{1\M}]-\E[Y_{0\M}]$ where $\M$ is defined to be the same as that observed distribution.

The identifying assumptions for $\text{GDE}(\M)$ are the same as those for CDE(5) in Table \ref{tab:CDE}, except that the service use score 5 is replaced with all values $m$ from the support of the distribution $\M$ (the range of variable service use observed in non-psychiatric patients). This range is covariate-dependent, e.g., it may be different for people on different health insurance plans, or for people who did versus did not have a checkup in the previous year.

While the assumptions here are more complex than those for $\text{CDE}(m)$ where $m$ is a single value, practical considerations of the conditional independence assumption tend to be similar -- seeking in broad terms exposure-mediator, exposure-outcome and mediator-outcome confounders. The positivity of relevant mediator values assumption, however, deserves attention. It requires that for any $(C,L)$ pattern, regardless of exposure condition, the observed range of mediator values covers the range given that $(C,L)$ pattern in the distribution $\M$. In the current example, since $\M$ is defined based on the distribution in the non-psychiatric population, this means that within $(C,L)$ levels, (i) the effective service use score range in bipolar patients in intervention and (ii) the corresponding range in bipolar patients in usual care both cover (iii) the range of this variable in non-psychiatric patients.

Under these assumptions, $\text{GDE}(\M)$ is identified by $\E_C\big[\E_{L\mid C,A=1}\big(\E_\M\big\{\E[Y|C,L,M,A\!=\!1]\big\}\big)\big]-\E_C\big[
\E_{L\mid C,A=0}\big(\E_\M\big\{\E[Y|C,L,M,A\!=\!0]\big\}\big)\big]$, a straightforward application of (\ref{RaM}).

\smallskip

A side note: While this $\text{GDE}(\M)$ is a useful contrast to consider, branch management notes that, as an approximation of an anticipated situation, it has an important limitation. If the branch's intervention is effective in helping bipolar patients with symptom management, one would expect that the effective service use distribution would not be exactly the same with or without the intervention. (A patient with better managed bipolar symptoms might benefit more from support, e.g., because they are more likely answer the calls of a case manager.) This is a general limitation of controlled/generalized direct effects. It is hard to match them to plausible situations where in both the exposed and unexposed conditions the mediator (a variable that naturally is affected by exposure) could be fixed to one value or set to the same distribution.

Generalized direct effects are just one of many  types of contrasts that involve this potential outcome type. Let us examine another simple example.

\subsection{Application 4: effect of a not yet implemented program}\label{app4}

\noindent After a round of consultation between headquarters and branches, it is decided that more research needs to be done before deciding whether to adopt the sweeping system change. One question is what would be its effect on health and well-being, assuming the above-mentioned result of eliminating the difference between psychiatric and non-psychiatric patients in terms of effective use of services for chronic medical problems. Our local branch decides to look into this potential effect for our population of bipolar patients.
For a rough answer, we use data from the usual care condition, and consider the contrast $\tau_1=\E[Y_{0\M}]-\E[Y_0]$, where $\M$ is defined to be the same as in Application 3. 

\begin{table}[h]
    \centering
    \caption{Identifying assumptions for $\tau_1$}
    \label{tab:tau1}
    \scriptsize
    \begin{tabular}{lllcc}
        &&& \multicolumn{2}{l}{Related to}  \\
        && Assumptions
        & $\E[Y_{0\M}]$ & $\E[Y_0]$
        \\\hline
        \multicolumn{2}{l}{Consistency}
        & $Y=Y_{0m}$ if $A=0,M=m$
        & \checkmark
        \\
        && $Y=Y_0$ if $A=0$
        && \checkmark
        \\\hline
        \multicolumn{2}{l}{Conditional independence:}
        \\
        & exposure-outcome
        & $A\independent Y_{0m}\mid C$
        & \checkmark
        \\
        && $A\independent Y_0\mid C$
        && \checkmark
        \\
        & mediator-outcome
        & $M\independent Y_{0m}\mid C,L,A=0$
        & \checkmark
        \\\hline
        \multicolumn{2}{l}{Positivity:}
        \\
        & exposure condition
        & $\P(A=0\mid C)>0$
        & \checkmark & \checkmark
        \\
        & mediator values
        & $\P(M=m\mid C,L,A=0)>0$
        & \checkmark
        \\\hline
        \multicolumn{2}{l}{Range of $m$ values}
        & support of distribution $\M$
        & \checkmark
        \\
        && ($C$-dependent)
        \\\hline
    \end{tabular}
\end{table}

Identification of $\tau_1$ requires identifying the means of $Y_{0\M}$ and $Y_0$. Table \ref{tab:tau1} collects all the required assumptions (from the relevant sections above, or relevant rows of Table \ref{tab:3x5}). These assumptions are arguably weaker than those for $\text{GDE}(\M)$, because for $\tau_1$ we do not need to identify $\E[Y_{1\M}]$. 

Under these assumptions, $\tau_1=\E_C\big[\E_{L\mid C,A=0}\big(\E_\M\big\{\E[Y|C,L,M,A\!=\!0]\big\}\big)\big]-\E_C\big\{\E[Y|C,A=0]\big\}$, where the first term is the same as the second term in the result for $\text{GDE}(\M)$. To help the reader easily spot this and several other connections among the identification results of the various effects considered in the illustrative example, we gather all those results in Table \ref{tab:app-id}. The table also shows simplified results in the special case with no intermediate confounders.

\smallskip

After examining data from and consulting with branches, and after serious consideration of logistics, costs and benefits, leadership drops the plan for the system change. This concludes a chapter of our story. In the next sections we will pay more attention to the intervention program for bipolar patients at our local branch.

\begin{table}[t]
    \caption{Identification results for the effects in the applications (top panel, see assumptions in relevant sections) and their simplication in the special case with no $L$ (bottom panel)}
    \label{tab:app-id}
    \centering
    \resizebox{.85\textwidth}{!}{%
    \begin{tabular}{cl}
        \textbf{Application} & \textbf{General case with intermediate confounders $L$}
        \\\hline
        \\[-.7em]
        1 & $\text{TE}=\colorbox{RubineRed!55}{$\E_C\big\{\E[Y\mid C,A=1]\big\}$}-\colorbox{cyan!50}{$\E_C\big\{\E[Y\mid C,A=0]\big\}$}$ 
        \\[.3em]\hline
        \\[-.7em]
        2 & $\text{CDE}(5)=\E_C\big(\E_{L\mid C,A=1}\big\{\E[Y\mid C,L,A=1,M=5]\big\}\big)-\E_C\big(\E_{L\mid C,A=0}\big\{\E[Y\mid C,L,A=0,M=5]\big\}\big)$
        \\[.3em]\hline
        \\[-.7em]
        3 & $\text{GDE}(\M)=\E_C\big[\E_{L\mid C,A=1}(\E_\M\{\E[Y|C,L,M,A\!=\!1]\})\big]-\colorbox{green!35}{$\E_C\big[\E_{L\mid C,A=0}(\E_\M\{\E[Y|C,L,M,A\!=\!0]\})\big]$}$
        \\[.3em]\hline
        \\[-.7em]
        4 & $\tau_1=\colorbox{green!35}{$\E_C\big[\E_{L\mid C,A=0}(\E_\M\{\E[Y|C,L,M,A\!=\!0]\})\big]$}-\colorbox{cyan!50}{$\E_C\big\{\E[Y\mid C,A=0]\big\}$}$
        \\[.3em]\hline
        \\[-.7em]
        5 & $\text{IDE}_0=\colorbox{yellow}{$\E_C\big[\E_{L\mid C,A=1}(\E_{\M_{0\mid C}}\{\E[Y\mid C,L,M,A=1]\})\big]$}-\E_C\big[\E_{L\mid C,A=0}(\E_{\M_{0\mid C}}\{\E[Y\mid C,L,M,A=0]\})\big]$,
        \\
        & ~$\text{IIE}_1=\E_C\big[\E_{L\mid C,A=1}\big(\E_{\M_{1\mid C}}\{\E[Y\mid C,L,M,A=1]\}\big)\big]-\colorbox{yellow}{$\E_C\big[\E_{L\mid C,A=1}(\E_{\M_{0\mid C}}\{\E[Y\mid C,L,M,A=1]\})\big]$}$
        \\
        & where $\M_{0\mid C}$ is identified as:~~$\P(\M^\text{draw}=m\mid C,L)=\P(M=m\mid C,A=0)$,
        \\
        & ~~~and $\M_{1\mid C}$ is identified as:~~$\P(\M^\text{draw}=m\mid C,L)=\P(M=m\mid C,A=1)$
        \\[.3em]\hline
        \\[-.7em]
        6 & $\tau_2=\colorbox{yellow}{$\E_C\big[\E_{L\mid C,A=1}(\E_{\M_{0\mid C}}\{\E[Y\mid C,L,M,A=1]\})\big]$}-\colorbox{cyan!50}{$\E_C\big\{\E[Y\mid C,A=0]\big\}$}$
        \\
        & where $\M_{0\mid C}$ is identified the same as in Application 5,
        \\[.5em]
        & $\tau_3=\E_C\big[\E_{L\mid C,A=1}(\E_{\M_{\mid C,0,L_1}}\{\E[Y\mid C,L,M,A=1]\})\big]-\colorbox{cyan!50}{$\E_C\big\{\E[Y\mid C,A=0]\big\}$}$
        \\
        & where $\M_{\mid C,0,L_1}$ is identified as:~ $\P(\M^\text{draw}=m\mid C,L_1=l)=\P(M=m\mid C,L=l,A=0)$
        \\[.3em]\hline
        \\[-.7em]
        7 & natural (in)direct effects are not identified
        \\[.3em]\hline\hline
        \\[-.7em]
        \textbf{Application} & \textbf{Special case with no intermediate confounders $L$}
        \\\hline
        \\[-.7em]
        1 & same as in the general case 
        \\[.3em]\hline
        \\[-.7em]
        2 & $\text{CDE}(5)=\E_C\big\{\E[Y\mid C,A=1,M=5]\big\}-\E_C\big\{\E[Y\mid C,A=0,M=5]\big\}$
        \\[.3em]\hline
        \\[-.7em]
        3 & $\text{GDE}(\M)=\E_C\big(\E_\M\big\{\E[Y|C,M,A\!=\!1]\big\}\big)-\colorbox{ForestGreen!40}{$\E_C\big(\E_\M\big\{\E[Y|C,M,A\!=\!0]\big\}\big)$}$
        \\[.3em]\hline
        \\[-.7em]
        4 & $\tau_1=\colorbox{ForestGreen!40}{$\E_C\big(\E_\M\big\{\E[Y|C,M,A\!=\!0]\big\}\big)$}-\colorbox{cyan!50}{$\E_C\big\{\E[Y\mid C,A=0]\big\}$}$
        \\[.3em]\hline
        \\[-.7em]
        5 & $\text{IDE}_0=\colorbox{Dandelion!40}{$\E_C\big(\E_{M\mid C,A=0}\big\{\E[Y\mid C,M,A=1]\big\}\big)$}-\colorbox{cyan!50}{$\E_C\big\{\E[Y\mid C,A=0]\big\}$},$
        \\
        & ~$\text{IIE}_1=\colorbox{RubineRed!55}{$\E_C\big\{\E[Y\mid C,A=1]\big\}$}-\colorbox{Dandelion!40}{$\E_C\big(\E_{M\mid C,A=0}\big\{\E[Y\mid C,M,A=1]\big\}\big)$}$
        \\[.3em]\hline
        \\[-.7em]
        6 & $\tau_2=\colorbox{Dandelion!40}{$\E_C\big(\E_{M\mid C,A=0}\big\{\E[Y\mid C,M,A=1]\big\}\big)$}-\colorbox{cyan!50}{$\E_C\big\{\E[Y\mid C,A=0]\big\}$}$,
        \\[.7em]
        & not applicable to $\tau_3$, because $\tau_3$ is defined specifically in the context with $L$
        \\[.3em]\hline
        \\[-.7em]
        7 & $\text{NDE}_0=\colorbox{Dandelion!40}{$\E_C\big(\E_{M\mid C,A=0}\big\{\E[Y\mid C,M,A=1]\big\}\big)$}-\colorbox{cyan!50}{$\E_C\big\{\E[Y\mid C,A=0]\big\}$},$
        \\
        & ~$\text{NIE}_1=\colorbox{RubineRed!55}{$\E_C\big\{\E[Y\mid C,A=1]\big\}$}-\colorbox{Dandelion!40}{$\E_C\big(\E_{M\mid C,A=0}\big\{\E[Y\mid C,M,A=1]\big\}\big)$}$
        \\[.3em]\hline
    \end{tabular}%
    }
\end{table}

\section{Identification of $\E[Y_{a\M}]$ where $\M$ is defined based on potential mediator distribution(s)}

\noindent This case inherits the same assumptions and the same result (\ref{RaM}) from the previous case. The only complication is that (\ref{RaM}) involves averaging over the distribution $\M$, which in the current case is not known a priori. This means the distribution $\M$ itself needs to be identified, which requires \textit{identification of the potential mediator distribution(s)} used in the definition of $\M$. This adds components to the three assumptions.

As a notation reminder, we use $M_{a^*}$ with a generic index $a^*$ to denote a potential mediator whose distribution informs the definition of the interventional mediator distribution $\M$. If $\M$ is informed by both a distribution of $M_1$ and a distribution of $M_0$, then both distributions need to be identified and we simply combine the assumptions.
The definition of the distribution $\M$ determines the form of the distribution of $M_{a^*}$ that needs to be identified. We consider three cases that may be of interest in applications (and that can serve as building blocks for more complex scenarios):
\begin{enumerate}[itemsep=0pt]
    \item[(i)] $\M$ is defined unconditionally to be the same as the marginal distribution of $M_{a^*}$;
    \item[(ii)] $\M$ is defined conditional on $C$ to be the same as the distribution of $M_{a^*}$ given $C$;
    \item[(iii)] $\M$ is defined conditional on $(C,L_a)$ to be the same distribution as that of $M_{a^*}$ given $(C,L_{a^*})$, precisely, $\P(\M^\text{draw}=m\mid C,L_a=l)=\P(M_{a^*}=m\mid C,L_{a^*}=l)$, where $\M^\text{draw}$ denotes a draw from the distribution $\M$.
\end{enumerate}

\subsection{Identifying first the relevant $M_{a^*}$ distribution(s) then the $\M$ distribution}

\noindent Identification of the $M_{a^*}$ marginal distribution and the conditional distribution given $C$ (cases (i) and (ii)), not surprisingly, requires assumptions similar to those that identify $\E[Y_a]$: consistency of $M_{a^*}$, conditional independence $A\independent M_{a^*}\mid C$, and positivity of exposure condition $a^*$. Identification of the distribution of $M_{a^*}$ given $(C,L_{a^*})$ (case (iii)) additionally requires consistency of $L_{a^*}$.%
\footnote{Strictly speaking, it also requires $A\independent(L_{a^*},M_{a^*})\mid C$, but this is implied by the assumption $A\independent M_{a^*}\mid C$ and the fact that the $C$ variables we are considering are confounders. The reasoning here is similar to that in footnote \ref{fn10}. Because $L_{a^*}$ is a cause of $M_{a^*}$, the confounders of the $A$-$M_{a^*}$ relationship contain the confounders of the $A$-$L_{a^*}$ relationship.}

Under these assumptions, the conditional distribution of $M_{a^*}$ is identified by the corresponding conditional distribution of the observed $M$ under exposure condition $a^*$: $\P(\textcolor{purple}{M_{a^*}}=m\mid C)=\P(M=m\mid C,A=a^*)$ for case (ii); and $\P(\textcolor{purple}{M_{a^*}}=m\mid C,L_{a^*}=l)=\P(M=m\mid C,L=l,A=a^*)$ for case (iii). The result for case (i) is obtained by averaging case (ii) result over the distribution of $C$, so $\P(\textcolor{purple}{M_{a^*}}=m)=\E_C[\P(M=m\mid C,A=a^*)]$.

Connecting from here to the distribution $\M$ is straightforward. The assumptions remain the same, except in case (iii), if $a^*\neq a$, the positivity assumption $\P(A=a^*\mid C)$ is replaced with $\P(A=a^*\mid C,L)$. This is to ensure that the range of covariate values that the distribution $\M$ conditions on is covered by the corresponding range in the relevant observed mediator distribution. The distribution $\M$ in the three cases is identified as follows:
\begin{alignat*}{3}
    &\text{case (i):} &&\P(\M^\text{draw}=m)&&=\E_C[\P(M=m\mid C,A=a^*)],
    \\
    &\text{case (ii):}&&\P(\M^\text{draw}=m\mid C)&&=\P(M=m\mid C,A=a^*),
    \\
    &\text{case (iii):}~~~&&\P(\M^\text{draw}=m\mid C,L_a=l)&&=\P(M=m\mid C,L=l,A=a^*).
\end{alignat*}

% Let us refer to this identified result for the distribution of $M_{a^*}$ as the \textit{informing} distribution, which now informs us about the distribution $\M$. This connection only requires that if $\M$ conditions on any covariates, the range of covariates conditioned on be covered by the corresponding range in the informing distribution. This does not apply to case (i). In case (ii) this requirement is already satisfied by the positivity assumption $\P(A=a^*\mid C)>0$. In case (iii), if $a^*=a$, this requirement is automatically met; if $a^*\neq a$, it can be seen as another form of positivity, $\P(A=a^*\mid C,L)>0$.

\bigskip

Now we are ready to assemble the full set of identifying assumptions for $\E[Y_{a\M}]$. Note that the following assumption statements all refer to $m$ values in the support of the distribution $\M$. What specific range this is for the three cases will be clarified at the end.

\subsection{Consistency assumption}

\noindent In all cases, this assumption includes  \textit{consistency of the potential outcome} $Y_{am}$ for all values $m$ in the support of the distribution $\M$, plus \textit{consistency of the potential mediator} $M_{a^*}$ (i.e., $M=M_{a^*}$ if $A=a^*$). In case (iii), \textit{consistency of $L_a$ and of $L_{a^*}$} is also assumed.

\subsection{Conditional independence assumption}

\noindent Like before, \textit{exposure-outcome} and \textit{mediator-outcome} conditional independence is assumed: for all relevant values $m$ in the support of the distribution $\M$,
\begin{align}
    A&\independent Y_{am}\mid C,\tag{I$_{a\M}$-\textsc{ay}}\label{IaM-ay}
    \\
    M&\independent Y_{am}\mid C,L,A=a.\tag{I$_{a\M}$-\textsc{my}}\label{IaM-my}
\end{align}
In addition \textit{exposure-mediator} conditional independence is assumed:
\begin{align}
    A\independent M_{a^*}\mid C.~~~~~~~~~~~~\tag{I$_{a\M}$-\textsc{am}}\label{IaM-am}
\end{align}
Unlike when the distribution $\M$ was known, here no subset of $C$ can replace $C$ in all three components of this conditional independence assumption.%
\footnote{The reason is that the subset $C^{LY}$ that satifies the first two components may leave out some exposure-mediator confounders, but the third component requires including all such confounders. Yet if we consider components one at a time, $C$ in the first two components can be replaced by $C^{LY}$, and $C$ in the third component can be replaced by the subset of variables that have (i) an arrow to $A$ and (ii) an arrow to $L$ or an arrow to $M$ or both.}

% A technical note: Strictly speaking, in case (iii), the third component of this assumption is $A\independent (L_{a^*},M_{a^*})\mid C$, but this distinction does not make a practical difference.%
% \footnote{The reasoning here is similar to that in footnote \ref{fn10}. Because $L_{a^*}$ is a cause of $M_{a^*}$, the confounders of the $A$-$M_{a^*}$ relationship contain the confounders of the $A$-$L_{a^*}$ relationship.}

\subsection{Positivity assumption}

\noindent Like before, \textit{positivity of exposure condition $a$} and \textit{positivity of relevant mediator values} are required, i.e., $\P(A=a\mid C)>0$ and $\P(M=m\mid C,L,A=1)>0$ for all $m$ values in the support of the distribution $\M$. 
In addition, positivity of exposure condition $a^*$ (for all levels of $C$) is required, i.e., $\P(A=a^*\mid C)>0$. In case (iii), if $a^*\neq a$, this is replaced by the stronger assumption of positivity of exposure condition $a^*$ for all levels of $(C,L)$, i.e., $\P(A=a^*\mid C,L)>0$.

\subsection{Identification result}

\noindent The identification result is of the same form as that for when $\M$ is a known distribution:
\begin{align}
    \E[\textcolor{purple}{Y_{a\M}}]=\E_C[\,\E_{L\mid C,A=a}(\,\E_{\M}\{\,\E[Y\mid C,L,M,A=a]\,\}\,)\,],\tag{\ref{RaM}}
\end{align}
except the distribution $\M$ here is not known but is identified as a function of the observed data distribution. In cases (ii) and (iii), this result could be written out in a simple format:
\begin{alignat*}{3}
    &\text{case (ii):}&&\E[\textcolor{purple}{Y_{a\M}}]&&=\E_C[\,\E_{L\mid C,A=a}(\,\E_{M\mid C,A=a^*}\{\,\E[Y\mid C,L,M,A=a]\,\}\,)\,],
    \\
    &\text{case (iii):}~~~&&\E[\textcolor{purple}{Y_{a\M}}]&&=\E_C[\,\E_{L\mid C,A=a}(\,\E_{M\mid C,L,A=a^*}\{\,\E[Y\mid C,L,M,A=a]\,\}\,)\,].
\end{alignat*}
For case (i), the expression%
\footnote{case (i): $\displaystyle\E[\textcolor{purple}{Y_{a\M}}]=\E_C\left\{\E_{L\mid C,A=a}\left[\int\E[Y\mid C,L,M=m,A=a]\,\E_C[\P(M=m\mid C,A=a^*)]\,\text{d}m\right]\right\}$.}
is complicated.

\subsection{The support of the distribution $\M$}

\noindent To be complete, we now clarify the support of the distribution $\M$ mentioned in the assumptions. In case (i), the support of the distribution $\M$ is the same as the support of variable $M$ given $A=a^*$. In case (ii), it is the same as the support of variable $M$ given $C,A=a^*$. In case (iii), it is the same as the support of variable $M$ given $C,A=a^*,L=l$ where $l$ is the actual value of $L_a$. These details will become more salient in  the applications.

\subsection{Application 5: interventional (in)direct effects}\label{app5}

\noindent Even though headquarters' plan was dropped, the debate about that plan raised branch management's interest in taking a closer look at the local intervention for bipolar patients, with special attention to the measure of effective service use. The interest is in quantifying how much of the intervention's effect is through this variable ($M$) and how much otherwise. This points to natural (in)direct effects (which decompose the total effect), but these effects are not identified in the current setting (this will be clear when we visit the fifth potential outcome type). Some suggest using interventional (in)direct effects as an approximation; others have reservations about this \cite[see arguments in][]{Nguyen2020}. Although there is no agreement, for the moment we consider identification of interventional (in)direct effects.

Consider a specific effect pair, $\text{IDE}_0=\E[Y_{1\M_{0\mid C}}]-\E[Y_{0\M_{0\mid C}}]$ and $\text{IIE}_1=\E[Y_{1\M_{1\mid C}}]-\E[Y_{1\M_{0\mid C}}]$, which involves three potential outcomes. $Y_{1\M_{1\mid C}}$ ($Y_{0\M_{0\mid C}}$) is the potential outcome in a hypothetical world where the patient receives the intervention (usual care), but the service use variable, instead of arising naturally and revealing the individual specific $M_1$ ($M_0$) value, is assigned a value drawn from the population distribution of $M_1$ ($M_0$) given $C$. 
$Y_{1\M_{0\mid C}}$ is the potential outcome in a hypothetical world where the patient receives the intervention but is assigned a mediator value drawn from the population distribution of $M_0$ given $C$. The identifying assumptions are collected in Table \ref{tab:IDEIIE}.

\begin{table}[h]
    \centering
    \caption{Identifying assumptions for $\textup{IDE}_0=\E[Y_{1\M_{0\mid C}}]-\E[Y_{0\M_{0\mid C}}]$ and $\textup{IIE}_1=\E[Y_{1\M_{1\mid C}}]-\E[Y_{1\M_{0\mid C}}]$}
    \label{tab:IDEIIE}
    \scriptsize
    \begin{tabular}{lllccc}
        &&& \multicolumn{3}{l}{Related to}
        \\
        && Assumptions
        & $\E[Y_{1\M_{1\mid C}}]$ 
        & $\E[Y_{1\M_{0\mid C}}]$
        & $\E[Y_{0\M_{0\mid C}}]$
        \\\hline
        \multicolumn{2}{l}{Consistency:}
        \\
        & potential outcome
        & $Y=Y_{1m}$ if $A=1,M=m$
        & \checkmark & \checkmark
        \\
        && $Y=Y_{0n}$ if $A=0,M=n$
        &&& \checkmark
        \\
        & potential mediator
        & $M=M_0$ if $A=0$
        && \checkmark & \checkmark
        \\\hline
        \multicolumn{2}{l}{Conditional independence:}
        \\
        & exposure-mediator
        & $A\independent M_1\mid C$
        & \checkmark
        \\
        && $A\independent M_0\mid C$
        && \checkmark & \checkmark
        \\
        & exposure-outcome
        & $A\independent Y_{1m}\mid C$
        & \checkmark & \checkmark
        \\
        && $A\independent Y_{0n}\mid C$
        &&& \checkmark
        \\
        & mediator-outcome
        & $M\independent Y_{1m}\mid C,L,A=1$
        & \checkmark & \checkmark
        \\
        && $M\independent Y_{0n}\mid C,L,A=0$
        &&& \checkmark
        \\\hline
        \multicolumn{2}{l}{Positivity:}
        \\
        & exposure condition
        & $\P(A=1\mid C)>0$
        & \checkmark & \checkmark
        \\
        && $\P(A=0\mid C)>0$
        && \checkmark & \checkmark
        \\
        & mediator values
        & $\P(M=m\mid C,L,A=1)>0$
        & \checkmark & \checkmark
        \\
        && $\P(M=n\mid C,L,A=0)>0$
        &&& \checkmark
        \\\hline
        \multicolumn{2}{l}{Ranges of $m$ and $n$}
        & $m$: the support of distribution
        \\
        && of $M$ given $C$
        & \checkmark* & \checkmark*
        \\
        && $n$: the support of distribution
        \\
        && of $M$ given $C,A=0$
        &&& \checkmark
        \\\hline
    \end{tabular}
    \caption*{\normalfont\scriptsize * $\E[Y_{1\M_{1\mid C}}]$ and $\E[Y_{1\M_{0\mid C}}]$ identifying assumptions involve $m$ values in the support of $M$ given $C,A=1$ and $m$ values in the support of $M$ given $C,A=0$, respectively. The combined range is the support of $M$ given $C$.}
\end{table}

While the statement of the conditional independence assumption here is complex, satisfying it in applications boils down to capturing all exposure-mediator and exposure-outcome confounders (in $C$), and capturing all mediator-outcome confounders within each exposure condition (in $C,L$). The positivity assumption, however, deserves more attention than it has been paid in the literature, as it may often be violated. 

Note that there is asymmetry in the ranges of mediator values for which the assumptions need to hold, which reflects the asymmetry of $Y_{1\M_{0\mid C}}$. We represent this asymmetry via separate expressions of $m$ and $n$ value ranges. The $m$ value range that defines the collection of potential outcomes $Y_{1m}$ is larger than the $n$ value range that defines the collection of potential outcomes $Y_{0n}$. The former is the support of the observed $M$ given $C$ (or the combined support of $M_1$ and $M_0$ given $C$); the latter is the support of $M$ given $C$ and $A=0$ (or the support of $M_0$ given $C$).

Consider the $n$-specific component $\P(M=n\mid C,L,A=0)>0$ for all $n$ values in the support of $M$ given $C,A=0$. To simplify reasoning, we condition on $C$ and $A=0$, and consider a subpopulation of patients in usual care that share the same values of $C$. Within such a subpopulation, this assumption means that the range of $M$ values does not depend on $L$, otherwise there are $L$ values for which the range of $M$ does not fully cover the support of $M$ in the subpopulation. In our example, if there is such a subpopulation (patients with a certain profile defined by baseline covariates $C$ who receive usual care) for whom the range of effective service use score ($M$) depends on the level of symptom management ($L$) (e.g., patients with poor symptom management have this score in the range of 0 to 3, while the full range for this subpopulation is 0 to 5), then this assumption is violated. The $m$-related component, $\P(M=m\mid C,L,A=1)>0$ for all $m$ values in the support of $M$ given $C$, is even more restrictive in that it requires not only that the range of $M$ given $C$ in the exposed not depend on $L$, but also that it cover the corresponding range in the unexposed. 

Putting the two components together, the \textit{positivity of relevant mediator values} assumption for identification of this effect pair means: within levels of $C$, (i) the range of $M$ if unexposed does not depend on $L$, (ii) the range of $M$ if exposed does not depend on $L$, and (iii) the latter covers the former. For any application, this  stringent positivity assumption should be checked against data.

In the special case with no intermediate confounders, then $L$ is an empty set, which substantially simplifies the positivity requirement. For example, if instead of service use we take symptom management to be the mediator of interest $M$, and define $\text{IDE}_0$ and $\text{IIE}_1$ the same way as above, except changing the mediator variable, then \textit{positivity of relevant mediator values} simply means that (iii) within levels of baseline covariates ($C$), the range of symptom management ($M$) under intervention covers the range under usual care.

% For the identification result of $\text{IDE}_0$ and $\text{IIE}_1$ (under the assumptions in Table \ref{tab:tau2}), see row 5 in the top panel of Table \ref{tab:app-id}. Row 5 in the bottom panel of the Table shows the expression this result simplifies to in the special case with no intermediate confounders.

Under the assumptions in Table \ref{tab:IDEIIE}, the identification result is:
\begin{align*}
    \text{IDE}_0&=\E_C\big[\E_{L\mid C,A=1}\big(\E_{\M_{0\mid C}}\big\{\E[Y\mid C,L,M,A=1]\big\}\big)\big]-\E_C\big[\E_{L\mid C,A=0}\big(\E_{\M_{0\mid C}}\big\{\E[Y\mid C,L,M,A=0]\big\}\big)\big],
    \\
    \text{IIE}_1&=\E_C\big[\E_{L\mid C,A=1}\big(\E_{\M_{1\mid C}}\big\{\E[Y\mid C,L,M,A=1]\big\}\big)\big]-\E_C\big[\E_{L\mid C,A=1}\big(\E_{\M_{0\mid C}}\big\{\E[Y\mid C,L,M,A=1]\big\}\big)\big],
\end{align*}
where the interventional mediator distributions $\M_{0\mid C}$ and $\M_{1\mid C}$ are identified as
$\P(\M^\text{draw}=m\mid C,L)=\P(M=m\mid C,A=0)$
and
$\P(\M^\text{draw}=m\mid C,L)=\P(M=m\mid C,A=1)$,
respectively (both independent of $L$ given $C$).
In the special case with no $L$ (see Table \ref{tab:app-id}, bottom panel), this simplifies and coincides with the result for natural (in)direct effects (in Application 7, section \ref{app7}).

\subsection{Application 6: effect of a modified intervention}\label{app6}

\noindent We continue to treat the service use variable as the mediator $M$. Now we have a different problem. Anticipating funding cuts we need to trim the intervention down to a lighter version, so a question is whether to remove the care-management component from the intervention and keep only the self-management component. With such a modified intervention, we expect that effective service use scores may be lower compared to levels under the original intervention, but would not be lower than levels under usual care.

What would be the effect on quality of life of a modified intervention removing care management? Our first answer is this effect could be conservatively approximated by $\tau_2=\E[Y_{1\M_{0\mid C}}]-\E[Y_0]$, where $Y_{1\M_{0\mid C}}$ is the potential outcome in a hypothetical world where the exposure is set to 1 (intervention) and the mediator is assigned a value drawn from the distribution of $M_0$ (effective service use distribution under usual care) given $C$. This assumes that the effective service use distribution under the modified intervention is the same as that under usual care. The identifying assumptions are collected in Table \ref{tab:tau2}. 

\begin{table}[h]
    \centering
    \caption{Identifying assumptions for $\tau_2=\E[Y_{1\M_{0\mid C}}]-\E[Y_0]$}
    \label{tab:tau2}
    \footnotesize
    \begin{tabular}{lllcc}
        &&& \multicolumn{2}{l}{Related to}
        \\
        && Assumptions
        & $\E[Y_{1\M_{0\mid C}}]$ & $\E[Y_0]$
        \\\hline
        \multicolumn{2}{l}{Consistency:}
        \\
        & potential outcome
        & $Y=Y_{1m}$ if $A=1,M=m$
        & \checkmark
        \\
        && $Y=Y_0$ if $A=0$
        && \checkmark
        \\
        & potential mediator
        & $M=M_0$ if $A=0$
        & \checkmark
        \\\hline
        \multicolumn{2}{l}{Conditional independence:}
        \\
        & exposure-mediator
        & $A\independent M_0\mid C$
        & \checkmark
        \\
        & exposure-outcome
        & $A\independent Y_{1m}\mid C$
        & \checkmark
        \\
        && $A\independent Y_0\mid C$
        && \checkmark
        \\
        & mediator-outcome
        & $M\independent Y_{1m}\mid C,L,A=1$
        & \checkmark
        \\\hline
        \multicolumn{2}{l}{Positivity:}
        \\
        & exposure condition
        & $\P(A=1\mid C)>0$
        & \checkmark
        \\
        && $\P(A=0\mid C)>0$
        & \checkmark & \checkmark
        \\
        & mediator values
        & $\P(M=m\mid C,L,A=1)>0$
        & \checkmark
        \\\hline
        \multicolumn{2}{l}{Range of $m$ values}
        & the support of distribution
        \\
        && of $M$ given $C,A=0$
        & \checkmark
        \\\hline
    \end{tabular}
\end{table}

As $\tau_2$ and $\text{IDE}_0$ share the same active intervention condition, but differ in the comparison condition, let us compare their identifying assumptions. There are differences in the consistency and conditional independence assumptions between the two contrasts, but these are not likely to matter in most applications. For example, with a rich enough collection of mediator-outcome confounders that we are willing to assume $M\independent Y_{1m}\mid C,L,A=1$ (which is required for both effects), it is likely that we are also willing to assume $M\independent Y_{0m}\mid C,L,A=0$ (which is additionally required by $\text{IDE}_0$), simply because there is a limit to how deeply we can realistically think about these assumptions. But again, the difference in the \textit{positivity of relevant mediator values} assumption has practical implications. This assumption, for $\tau_2$, is simply that within levels of baseline covariates ($C$), the effective service use ($M$) range under intervention, regardless of symptom management ($L$) value, has to cover the full range under usual care. For $\text{IDE}_0$, however, the assumption also requires that within levels of $C$, the $M$ range if unexposed does not depend on $L$ values. This means positivity is more likely to hold for $\tau_2$ than for $\text{IDE}_0$.

Under the assumptions in Table \ref{tab:tau2}, $\tau_2$ is identified by $\E_C\big[\E_{L\mid C,A=1}\big(\E_{\M_{0\mid C}}\big\{\E[Y\mid C,L,M,A=1]\big\}\big)\big]-\E_C\big\{\E[Y\mid C,A=0]\big\}$,
where the first term is the same as the first term in the result for $\text{IDE}_0$, but the second term is simpler (see Table \ref{tab:app-id}). In the special case with no $L$, the two results simplify and coincide with each other, and coincide with the result for $\text{NDE}_0$.

\smallskip

As $\tau_2$ is a conservative approximation of the effect of the modified intervention, we also consider a closer approximation that does not fix the mediator distribution at $\M_{0\mid C}$. The rationale is that improvement in symptom management (which results from the self-management component of the intervention) may itself lead to more effective service use. We thus consider $\tau_3=\E[Y_{1\M_{\mid C,0,L_1}}]-\E[Y_0]$, 
where $Y_{1\M_{\mid C,0,L_1}}$ is the potential outcome where everything occurs as if in the original intervention condition, except that the effective service use variable is shifted to a distribution $\M_{\mid C,0,L_1}$ that conditions on covariate values $(C,L_1)$ but is defined to be the same as the distribution of $M_0$ given $(C,L_0)$. That is, $\P(\M_{\mid C,0,L_1}^\text{draw}=m\mid C,L_1=l)=\P(M_0=m\mid C,L_0=l)$. Intuitively, this distribution allows the change in symptom management to influence effective service use. 

\begin{table}[h]
    \centering
    \caption{Identifying assumptions for $\tau_3=\E[Y_{1\M_{\mid C,0,L_1}}]-\E[Y_0]$}
    \label{tab:tau3}
    \footnotesize
    \begin{tabular}{lllcc}
        &&& \multicolumn{2}{l}{Related to}
        \\
        && Assumptions
        & $\E[Y_{1\M_{\mid C,0,L_1}}]$ & $\E[Y_0]$
        \\\hline
        \multicolumn{2}{l}{Consistency:}
        \\
        & potential outcomes
        & $Y=Y_{1m}$ if $A=1,M=m$
        & \checkmark
        \\
        && $Y=Y_0$ if $A=0$
        && \checkmark
        \\
        & potential mediator
        & $M=M_0$ if $A=0$
        & \checkmark
        \\
        & potential int. confnders
        & $L=AL_1+(1-A)L_0$
        & \checkmark
        \\\hline
        \multicolumn{2}{l}{Conditional independence:}
        \\
        & exposure-mediator
        & $A\independent M_0\mid C$
        & \checkmark
        \\
        & exposure-outcome
        & $A\independent Y_{1m}\mid C$
        & \checkmark
        \\
        && $A\independent Y_0\mid C$
        && \checkmark
        \\
        & mediator-outcome
        & $M\independent Y_{1m}\mid C,L,A=1$
        & \checkmark
        \\\hline
        \multicolumn{2}{l}{Positivity:}
        \\
        & exposure condition
        & $\P(A=1\mid C)>0$
        & \checkmark
        \\
        && $\P(A=0\mid C,L)>0$
        & \checkmark & \checkmark*
        \\
        & mediator values
        & $\P(M=m\mid C,L,A=1)>0$
        & \checkmark
        \\\hline
        \multicolumn{2}{l}{Range of $m$ values}
        & the support of distribution of
        \\
        && $M$ given $C,L=l,A=0$, where
        & \checkmark
        \\
        && $l$ is the actual value of $L_1$
        \\\hline
    \end{tabular}
\end{table}

The identifying assumptions for $\tau_3$ are shown in Table \ref{tab:tau3}. Comparing to $\tau_2$, there are two differences in the positivity assumption. The \textit{positivity of the unexposed condition} assumption is more restrictive for $\tau_3$ than for $\tau_2$: for $\tau_3$ it requires that for any realized values of $(C,L)$ combined, the probability of receiving usual care is positive; for $\tau_2$ it only requires this for all $C$ values. On the other hand, the \textit{positivity of relevant mediator values} assumption is more restrictive for $\tau_2$: for $\tau_2$ it requires that within levels of $C$, the $M$ range under intervention for any $L$ value covers the full $M$ range under usual care; for $\tau_3$ it requires that within levels of $(C,L)$, the $M$ range under intervention covers the corresponding range under usual care. Footnote 
\footnote{Consider a subpopulation homogeneous in $C$. For simplicity, symptom management ($L$) has two levels, high and low. First, suppose that the service use ($M$) range under intervention is 0 to 3 if low $L$ and 0 to 5 if high $L$, and the $M$ range in usual care is 0 to 3 for both low and high $L$; in this case the \textit{positivity of relevant mediator values} assumption holds for both $\tau_2$ and $\tau_3$ because both of the ranges under intervention cover the full range under usual care. Second, if instead the $M$ range in usual care is 0 to 3 if low $L$ and 0 to 4 if high $L$, the assumption holds for $\tau_3$ but not $\tau_2$, because the intervention range covers the usual care range in low and high $L$ strata, but one stratum-specific intervention range does not cover the full usual care range. Third, if one (or both) of the ranges under intervention is shifted up and does not include value 0, then the assumption does not hold either for $\tau_2$ or $\tau_3$.}
clarifies this using numeric values.

Under the assumptions in Table \ref{tab:tau3}, $\tau_3=\E_C\big[\E_{L\mid C,A=1}\big(\E_{\M_{\mid C,0,L_1}}\big\{\E[Y\mid C,L,M,A=1]\big\}\big)\big]-\E_C\big\{\E[Y\mid C,A=0]\big\}$, where the interventional mediator distribution $\M_{\mid C,0,L_1}$ is identified as $\P(\M^\text{draw}=m\mid C,L_1=l)=\P(M=m\mid C,L=l,A=0)$.

\subsection{Recap so far}

\noindent We have seen that as the condition that defines the potential outcome gets more complex, the assumptions required for identifying its mean get more complicated. The conditional independence assumption evolved from a single component (\textit{exposure-outcome} conditional independence) for $\E[Y_a]$ identification, to pairs of components (\textit{exposure-outcome} and \textit{mediator-outcome}) for identification of $\E[Y_{am}]$ and of $\E[Y_{a\M}]$ where $\M$ is a known distribution, to a trio of components (\textit{exposure-mediator}, \textit{exposure-outcome} and \textit{mediator-outcome}) for identification of $\E[Y_{a\M}]$ where $\M$ is defined based on potential mediator distribution(s). As we shall see, the mean of the cross-world potential outcome requires the most complex conditional independence assumption.

But before diving into that last one, it helps to pause and take a high-level view of the types of reasoning we have engaged in so far, and consolidate two key moves we use. The first move, which we have used repeatedly, is \textit{going from whole to part} (or vice versa), made possible by conditional independence. The gist is that within levels of \textit{observed} covariates, when the \textit{observed} $A$ (or $M$) is independent of a \textit{potential} variable (mediator or outcome), that allows us to learn about the overall distribution (and mean) of that potential variable just from the values observed in a relevant subset of individuals. For example, within levels of $C$, we learn about the $Y_a$ distribution from the observed $Y$ of individuals with $A=a$ under (\ref{Ia}), and about the $M_{a^*}$ distribution from the observed $M$ of individuals with $A=a^*$ under (\ref{IaM-am}). Within levels of $C,L$, we learn about the distribution of $Y_{am}$ from individuals with $A=a,M=m$ under (\ref{Iam-ay}) and (\ref{Iam-my}) combined. 
The second move, which we use when considering a condition where the mediator is set to a certain distribution rather than is let to occur naturally, is \textit{swapping of mediator distributions}. When targeting the mean of $Y_{a\mathcal{M}}$, going from whole to part means we only need to consider the outcome in those with $A=a$. But this only gets us halfway there, because each outcome follows a mediator value, so the mean outcome in those with $A=a$ is tied to the corresponding observed mediator distribution. We need to swap this distribution out for the interventional mediator distribution $\mathcal{M}$. This is done above by averaging over the distribution $\mathcal{M}$ instead of over the observed mediator distribution -- see (\ref{RaM}).
Equipped with these two moves, we now tackle the cross-world potential outcome.

\section{Identification of $\E[Y_{aM_{a'}}]$ where $a\neq a'$}

\noindent It is important to note that this potential outcome is fundamentally different from the ones considered so far, in that it belongs in a completely counterfactual world where the mediator is set, not to a fixed value for all, not to a draw from a certain distribution, but to the specific value that the individual would experience under a different exposure. This cross-world potential outcome is unobservable. Yet its distribution (and mean) is identified under a set of assumptions. Let us now build some intuitive appreciation for the assumptions. (Interested readers are referred to the Appendix for precise proof, and to the relevant literature for extensive discussion of this potential outcome, e.g., \citeauthor{Pearl2001}, \citeyear{Pearl2001}; \citeauthor{VanderWeele2009}, \citeyear{VanderWeele2009}; \citeauthor{Imai2010}, \citeyear{Imai2010}; \citeauthor{Pearl2012}, \citeyear{Pearl2012}; \citeauthor{Robins2003}, \citeyear{Robins2003}.)

For clarity, let us sketch a picture of this hypothetical world. In this world, at first things happen naturally, including the pre-exposure covariates $C$. Then exposure is set to $a$. After that everything follows naturally from $a$ for a while; any intermediate confounder $L$ reveals the individual's $L_a$. Then right at the point where the mediator $M$ is about to realize (as $M_a$), it is magically intervened upon and set to the individual's $M_{a'}$ value. After that again everything, including the outcome, follows naturally. The outcome $Y_{aM_{a'}}$ arises based on the combination of $(C,a,L_a,M_{a'})$ where $L_a$ and $M_{a'}$ are from different worlds.

We need to somehow connect the mean of this unobservable $Y_{aM_{a'}}$ to observable data. First, we use the \textit{whole-to-part} move to narrow down to considering this potential outcome only among those in the $A=a$ condition. By conditioning on $C$ and invoking an \textit{exposure-outcome} conditional independence assumption similar to (\ref{IaM-ay}), within levels of $C$, we replace the mean of $Y_{aM_{a'}}$ with its mean among those who experience $A=a$. That is, $\E[Y_{aM_{a'}}\mid C]=\E[Y_{aM_{a'}}\mid C,A=a]$. This move does not make $Y_{aM_{a'}}$ any more observable (it is not) but matches the exposure condition; the mediator remains mismatched.

Next, consider only those in the $A=a$ condition. For each of these individuals, the observed mediator is $M_a$ (under consistency). Ideally we want to swap this $M_a$ value for the individual's $M_{a'}$ value in order to obtain knowledge about the target potential outcome $Y_{aM_{a'}}$, but unfortunately $M_{a'}$ is unobserved. Our strategy is to use as a proxy for $M_{a'}$ a distribution (to which $M_{a'}$ belongs) that captures all the information about $M_{a'}$ that is relevant for the purpose of identifying $\E[Y_{aM_{a'}}]$, and try to identify that distribution. Using the proxy distribution, we could then apply the \textit{swapping of mediator distributions} move to identify the target potential outcome mean. But what should be the proxy distribution? It turns out that to capture all the relevant information about $M_{a'}$, it has to be a distribution of $M_{a'}$ that conditions on a set of \textit{observed} variables that removes the confounding of the relationship between $M_{a'}$ and $Y_{aM_{a'}}$. 
% And it needs to be identified.

\begin{figure}[h]
\caption{Examining the common causes of $M_{a'}$ and $Y_{aM_{a'}}$}\label{fig:crossworldconfounding}
    \begin{center}
    \begin{tikzpicture}[
        box/.style={rectangle, minimum size=5mm}
    ]
        \node[box]  (C)    {$C$};
        \node[box]  (L1)  [right=of C, yshift=10mm] {$L_a$};
        \node[box]  (L0) [right=of C, yshift=-10mm] {$L_{a'}$};
        \node[box]  (M0)  [right=of L0] {$M_{a'}$};
        \node[box] (Y) [right=of M0, yshift=10mm] {$Y_{aM_{a'}}$};
        
        \node[box]  (U) [right=of C, xshift=6mm, yshift=-2mm] {\footnotesize $U_L$};
        
        \draw[->] (C) -- (L1);
        \draw[->] (C) -- (L0);
        \draw[->] (C) -- (Y);
        \draw[->] (C) .. controls +(down:15mm) and +(down:10mm)  .. (M0);
        \draw[->] (L1) -- (Y);
        \draw[->] (L0) -- (M0);
        \draw[->] (M0) -- (Y);
        \draw[->] (U) -- (L1);
        \draw[->] (U) -- (L0);
        
        \node[box]  (Cb) [right=of Y, xshift=10mm]    {$C$};
        \node[box]  (M0b) [right=of Cb, yshift=-10mm] {$M_{a'}$};
        \node[box] (Yb) [right=of M0b, yshift=10mm] {$Y_{aM_{a'}}$};
        
        \draw[->] (Cb) -- (M0b);
        \draw[->] (Cb) -- (Yb);
        \draw[->] (M0b) -- (Yb);
        
        \node[box] (a) [below=of L0, xshift=6mm, yshift=2mm]  {a. The general case};
        \node[box] (b) [right=of a, xshift=24mm] {b. The special case with no $L$};
        
    \end{tikzpicture}
    \end{center}
\end{figure}
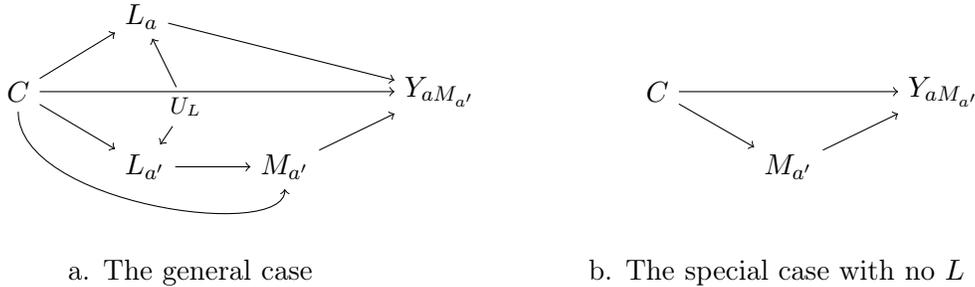

In the special case with no intermediate confounders, $C$ is the only common cause of $M_{a'}$ and $Y_{aM_{a'}}$: $C$ causes $M_{a'}$ in the world with exposure $a'$, and causes $Y_{aM_{a'}}$ directly in the hypothetical world we are considering. This simple confounding structure is shown in Fig. \ref{fig:crossworldconfounding}b.
The appropriate proxy for $M_{a'}$ in those with $A=a$ is thus the distribution of $M_{a'}$ given $C,A=a$. Although $M_{a'}$ is not observed for those with $A=a$, under the same \textit{exposure-mediator} conditional independence assumption as (\ref{IaM-am}) (except replacing $M_{a^*}$ with $M_{a'}$), this distribution is identified to be equal to the distribution of the observed $M$ given $C,A=a'$. This means that, assuming all the other assumptions hold, the result for $\E[Y_{aM_{a'}}]$ is a simple adaptation of the result for $\E[Y_{a\mathcal{M}}]$, replacing $\mathcal{M}$ with this proxy distribution and removing $L$.

In the general case, there are likely intermediate confounders. Now the relationship of $M_{a'}$ and $Y_{aM_{a'}}$ is confounded not only by $C$ but also by the unique cause $U_L$ of $L$: $U_L$ causes $M_{a'}$ through $L_{a'}$, and also causes $Y_{aM_{a'}}$ through $L_a$ (see Fig. \ref{fig:crossworldconfounding}a).
% due to the connection $M_{a'}\leftarrow L_{a'}\leftarrow U_L\to L_a\to Y_{aM_{a'}}$ ($U_L$ has a causal influence on $M_{a'}$ via $L_{a'}$, and a causal influence on $Y_{aM_{a'}}$ via $L_a$). 
This means that the set of variables that a proxy distribution (if one exists) conditions on must include not only $C$ but also at least one of the three variables $U_L,L_a,L_{a'}$ (to remove confounding by $U_L$). Among those with $A=a$ (whose outcomes we are counting on using to learn about the target potential outcome mean), only $L_a$ is observed. This suggests using the distribution of $M_{a'}$ given $C,L_a,A=a$ as the proxy distribution. Unfortunately, any distribution of $M_{a'}$ that conditions on the other-world $L_a$ is unidentified. Hence $\E[Y_{aM_{a'}}]$ is unidentified.
Consequently, identification of $\E[Y_{aM_{a'}}]$ requires that there are no intermediate confounders.

\subsection{Consistency assumption}

\noindent Similar to the previous case, this assumption here includes the regular (i) consistency of potential outcomes, $Y=Y_{am}$ if $A=a,M=m$ for $m$ in the support of the observed $M$ given $C,A=a'$; and (ii) consistency of the potential mediator, $M=M_{a'}$ if $A=a'$. In addition, it includes (iii) consistency of the cross-world potential outcome, $Y_{aM_{a'}}=Y_{am}$ if $M_{a'}=m$. The latter belongs in a different category that connect different types of potential variables, rather than connecting potential to observed variables.

\subsection{Conditional independence assumption}

\noindent To show how this assumption compares to the one for the previous potential outcome type, we present it in four elements as follows.
\begin{equation}
    A\independent M_{a'}\mid C,\tag{I$_{aM_{a'}}$-\textsc{am}}\label{Iaa'-am}
\end{equation}
and for all values $m$ in the support of $M$ given $C,A=a'$,
\begin{align}
    A&\independent Y_{am}\mid C,\tag{I$_{aM_{a'}}$-\textsc{ay}}\label{Iaa'-ay}
    \\
    M&\independent Y_{am}\mid C,A=a,\tag{I$_{aM_{a'}}$-\textsc{my1}}\label{Iaa'-my1}
    \\
    M_{a'}&\independent Y_{am}\mid C.\tag{I$_{aM_{a'}}$-\textsc{my2}}\label{Iaa'-my2}
\end{align}

The first three elements are similar to the assumption from the previous potential outcome type, except replacing $a^*$ with $a'$ and removing $L$ (since we require an empty $L$).%
\footnote{Strictly speaking, (\ref{Iaa'-my1}) can be relaxed to condition on some $L$ in addition to $C$, but it is required that (\ref{Iaa'-my2}) holds conditional on $C$ only. This applies to the very special case where there exist variables $L$ that are influenced by $A$ that influence the mediator and outcome under exposure $a$, but do not influence mediator under exposure $a'$; this amounts to removing the arrow from $L_{a'}$ to $M_{a'}$ in Fig. \ref{fig:crossworldconfounding}a.}
The key difference (from all the previous sets of assumptions) is the fourth element, which says that conditioning on $C$ is sufficient to remove confounding between $M_{a'}$ and $Y_{aM_{a'}}$.
% \footnote{This rules out special situations where (\ref{Iaa'-my1}) is met but not (\ref{Iaa'-my2}), e.g., when a variable influenced by exposure influences the outcome but not the mediator under exposure $a$, but influences the mediator under exposure $a'$.\label{footnote:rarecase}}
This fourth element is well known as the \textit{cross-world independence assumption}.

% To aid memory, the third and fourth elements, both concerning \textit{mediator-outcome} conditional independence, could be replaced by a single simple statement (technically a stronger assumption):
% \begin{align}
%     M_1,M_0&\independent Y_{am}\mid C,A,\tag{I$_{aM_{a'}}$-\textsc{my}}\label{Iaa'-my}
% \end{align}
% which says that within levels of $C$ and exposure conditions, both potential mediators are independent of $Y_{am}$.

\subsection{Positivity assumption}

\noindent This assumption includes \textit{positivity of both exposure conditions}, $0<\P(A=1\mid C)<1$; and \textit{positivity of relevant mediator values}, $\P(M=m\mid C,A=a)>0$ for all $m$ values in the support of $M$ given $C,A=a'$.

\subsection{Identification result}

\noindent The identification result is a rather simple triple expectation:

\begin{equation}
    \E[\textcolor{purple}{Y_{aM_{a'}}}]=\E_C(\,\E_{M\mid C,A=a'}\{\,\E[Y\mid C,M,A=a]\,\}\,).\tag{R$_{aM_{a'}}$}\label{Raa'}
\end{equation}

\medskip

Here when moving from the inner to the middle expectation in this result, we have swapped out (i) the mediator distribution in those whose outcome data we use for (ii) the mediator distribution in those with $A=a'$ within levels of $C$, which identifies (iii) the distribution of $M_{a'}$ given $C,A=a$, which is the proxy distribution.

It is noteworthy that (\ref{Raa'}) bears some resemblance to the identification result of $\E[Y_{a\M}]$ where the distribution $\M$ is defined to be the same as the distribution of $M_{a'}$ given $C$; the difference is that that previous result conditions on $L$ in the inner expectation and averages over $L$ in a middle expectation. In the special case with no intermediate confounders, that result reduces exactly to the current result, which is not surprising given that the proxy distribution we use above is equivalent to that definition of the distribution $\M$. Note, though, that the assumptions that lead from two different potential outcome means to the same identification result are not the same; we return to this point shortly.

\subsection{Application 7: a pair of natural (in)direct effects}\label{app7}

\noindent The cross-world potential outcome type is involved in the natural (in)direct effects. Here we consider one pair of these effects, which results from using $Y_{1M_0}$ to split the total effect:
$$\text{TE}=\overbrace{\E[Y_1]-\E[Y_{1M_0}]}^{\textstyle\text{NIE}_1}-\overbrace{\E[Y_{1M_0}]-\E[Y_0]}^{\textstyle\text{NDE}_0}.$$
These contrasts are interpreted as (in)direct effects because we could re-express $Y_1$ as $Y_{1M_1}$ and $Y_0$ as $Y_{0M_0}$, so one contrast is the effect of switching exposure while the mediator is fixed, and the other is the effect of switching mediator value while exposure is fixed. The re-expression $Y_a=Y_{aM_a}$ is actually another consistency assumption that connects the two potential outcomes, commonly known as the \textit{composition} assumption \citep{VanderWeele2009}. This assumption is only needed for the interpretation.

\begin{table}[h]
    \centering
    \caption{Identifying assumptions for $\textup{NDE}_0$ and $\textup{NIE}_1$}
    \label{tab:NDENIE}
    \scriptsize
    \begin{tabular}{lllccc}
        &&& \multicolumn{2}{l}{Related to}  
        \\
        && Assumptions 
        & $\E[Y_1]$ and $\E[Y_0]$ & $\E[Y_{1M_0}]$ 
        \\\hline
        \multicolumn{2}{l}{Consistency:}
        \\
        & potential outcomes
        & $Y=AY_1+(1-A)Y_0$
        & \checkmark
        \\
        && $Y_a=Y_{aM_a}$ for $a=0,1$
        \\
        && $Y=Y_{1m}$ if $A=1,M=m$
        && \checkmark
        \\
        && $Y_{aM_{a'}}=Y_{am}$ if $M_{a'}=m$
        && \checkmark
        \\
        & potential mediator
        & $M=M_0$ if $A=0$
        && \checkmark
        \\\hline
        \multicolumn{2}{l}{Conditional independence:}
        \\
        & exposure-outcome
        & $A\independent Y_a\mid C$ for $A=0,1$
        & \checkmark
        \\
        && $A\independent Y_{1m}\mid C$
        && \checkmark
        \\
        & exposure-mediator
        & $A\independent M_0\mid C$
        && \checkmark
        \\
        & mediator-outcome 
        & $M\independent Y_{1m}\mid C,A=1$
        && \checkmark
        \\
        && $M_0\independent Y_{1m}\mid C$
        && \checkmark
        \\\hline
        \multicolumn{2}{l}{Positivity:}
        \\
        & exposure condition
        & $0<\P(A=0\mid C)<1$
        & \checkmark & \checkmark
        \\
        & mediator values
        & $\P(M=m\mid C,A=1)>0$
        && \checkmark
        \\\hline
        \multicolumn{2}{l}{Range of $m$ values}
        & the support of distribution
        \\
        && of $M$ given $C,A=0$
        && \checkmark
        \\\hline
    \end{tabular}
\end{table}

Identification of this effect pair requires identifying the means of $Y_0$, $Y_1$ and $Y_{1M_0}$. The required assumptions are collected in Table \ref{tab:NDENIE}.

Before looking more closely at these assumptions, we note that for our example, with the service use variable as the mediator ($M$), natural (in)direct effects are not identified due to the presence of the intermediate confounder symptom management ($L$); this violates the key cross-world independence assumption required to identify $\E[Y_{1M_0}]$. 

While natural (in)direct effects are not identified, recall from Application 5 that if the assumptions in Table \ref{tab:IDEIIE} hold, interventional (in)direct effects are identified. This is perhaps a common situation, as the absence of intermediate confounders may be a special situation; it has been noted that unless the mediator comes closely in time after the exposure, there are likely intermediate confounders \citep{Vansteelandt2012}. In such a situation, it might be tempting to want to switch the estimand from natural to interventional (in)direct effects (as in our story in Application 5), but then that effectively changes the question being asked. An alternative is to opt for estimating bounds on natural (in)direct effects \citep{Miles2017} or to do a sensitivity analysis on the unidentified cross-world associations \citep{Daniel2015}.

Let us put our bipolar intervention example with the intermediate confounder problem aside and examine the assumptions in Table \ref{tab:NDENIE} as generic assumptions. We note that these assumptions are asymmetric, which means they are weaker than the symmetric assumptions often found in the literature. While the symmetric assumptions allow identification of both pairs of natural (in)direct effects, it seems that often researchers are interested in only one of the two pairs, which means only the asymmetric assumptions relevant to the pair are required. Admittedly it may be hard to think in practical terms about when the conditional independence assumptions may hold for one but not the other pair of natural (in)direct effects. But this is more clear with the \textit{positivity of mediator values} assumption. Its symmetric statement in the literature implies that within levels of $C$, the mediator range is the same between the two exposure conditions, which is unnecessarily restrictive if we are interested in only one pair of effects. For the current pair of effects, it is only required that within levels of $C$, the mediator range in the exposed condition covers the mediator range in the unexposed condition. For the other pair of effects, the opposite is required. The practical implication is that in some cases, one pair of natural (in)direct effects may be identified but the other is not due to positivity violation.

% For example, we need $C$ to capture all common causes of $A$ and $M_0$ but not necessarily cover all common causes of $A$ and $M_1$. This may be relevant if there is a known cause of $A$ that is not measured in the study (therefore not included in $C$) that influences variation of the mediator in the exposed condition but not in the unexposed condition, as that does not violate $A\independent M_0\mid C$. The positivity of mediator values assumption requires that within levels of $C$, the mediator range in the exposed condition covers the mediator range in the unexposed condition; the opposite is not required. This is important to note, because the symmetric statement of positivity often found in the literature, which implies that within levels of $C$ the mediator range is the same between the two exposure conditions, may not hold in applications and is unnecessarily restrictive.

% Coming back to our example, treating use of care as the mediator, the presence of the intermediate confounder symptom care rules out identification of the natural (in)direct effects. If we treat symptom management as the mediator of interest (calling it $M$), however, there is (we believe) no intermediate confounders. This means if we are willing to make all the identifying assumptions above, the natural (in)direct effects corresponding to this mediator variable are identified.

Under the assumptions in Table \ref{tab:NDENIE}, the current pair of natural (in)direct effects are identified as
\begin{align*}
    \text{NDE}_0&=\E_C\big(\E_{M\mid C,A=0}\big\{\E[Y\mid C,M,A=1]\big\}\big)-\E_C\big\{\E[Y\mid C,A=0]\big\},
    \\
    \text{NIE}_1&=\E_C\big\{\E[Y\mid C,A=1]\big\}-\E_C\big(\E_{M\mid C,A=0}\big\{\E[Y\mid C,M,A=1]\big\}\big).
\end{align*}

\subsubsection{A foot note on extension of natural effects in the case with $L$}

The presence of intermediate confounders $L$ means that the natural (in)direct effects are unidentified. In this case, an alternative is to treat $L$ and $M$ both as mediators and target path-specific effects \citep[see e.g.,][]{VanderWeele2014a}. This would take us to the general topic of multiple mediators analysis, which is outside the scope of the current paper. Here we just note a few key points without explication. Path-specific effects are extensions of natural (in)direct effects for the multiple causally ordered mediators case. Definition of these effects requires a different kind of nested potential outcome, $Y_{aL_{a'}M_{a''L_{a'}}}$ (for the two mediators case). For example, one decomposition of the total effect is into three components: a direct effect ($\E[Y_{1L_0M_{0L_0}}]-\E[Y_0]$), an effect through the first mediator ($\E[Y_{1L_1M_{0L_1}}]-\E[Y_{1L_0M_{0L_0}}]$) and an effect through the second but not the first mediator ($\E[Y_1]-\E[Y_{1L_1M_{0L_1}}]$). Roughly speaking, identification of these three effects requires that there are no unobserved exposure-mediator, exposure-outcome, mediator-mediator ($L$-$M$) and mediator-outcome confounders, and that there are no exposure-induced (observed or unobserved) confounders of the relationship between $(L,M)$ and $Y$, plus relevant consistency and positivity assumptions.

\section{Concluding remarks}

\noindent We have shown that identification of a wide range of causal effects in the single mediator case boils down to identification of the mean (distribution) of potential outcomes of five types, and how the assumptions required are connected, getting more complex only as the condition that defines the potential outcome gets more complex. We provide Table \ref{tab:3x5} as a menu the substantive researcher can use to assemble identifying assumptions for their target causal estimand. We demonstrate the plausibility consideration of such assumptions for several estimands of common interest through an illustrative example. 

We recommend using this paper alongside the companion ``estimands'' paper \cite{Nguyen2020}. The combination of the two papers aim to help the applied researcher first flexibly define causal mediation effects to match their research question and then to assess the effects' identifiability.

This paper did not cover more complex cases such as multiple causally ordered or unordered mediators and repeated exposure and/or mediator over a longitudinal process. Each of these settings comes with a range of effect definitions and identification strategies. The same kind of exercises we conducted in the simple case (connecting effect definitions to real-world research questions and systematic examination of their identification assumptions) is highly recommended for these more complex cases -- for the purpose of making advanced methods more accessible and meaningful to applied researchers, facilitating their appropriate use and promoting quality research.

\section*{Acknowledgements}

TQN's and EAS's work on this article was supported by grant R01MH115487 (PI Stuart), and IS's was supported by grant T32MH122357 (PI Stuart) from the National Institute of Mental Health. All viewpoints and errors belong to the authors. This work has benefited from the helpful feedback we received from several reviewers and from students who participated in our mediation analysis course at Johns Hopkins in 2021 and 2022.

\bibliography{refs}

\appendix

\section*{Appendix}
\allowdisplaybreaks
\singlespace

\noindent We derive the identification results for the five potential outcome means based on the corresponding consistency, conditional independence and positivity assumptions.

\subsection*{Identification of $\E[Y_a]$}

\begin{align*}
    \E[Y_a]
    &=\E\{\E[Y_a\mid C]\} & \text{\scriptsize(iterated expectation)}
    \\
    &=\E\{\E[Y_a\mid C,A=a]\} & \text{\scriptsize(conditional independence $A\independent Y_a\mid C$)}
    \\
    &=\E\{\E[Y\mid C,A=a]\}. & \text{\scriptsize(consistency and positivity)}
\end{align*}
In the main text, for an expression with multiple layers of expectations, we adopt an index notation for all the outer layers (excluding the innermost one). This result is thus stated,
$$\E[Y_a]=\E_C\{\E[Y\mid C,A=a]\}.$$

\subsection*{Identification of $\E[Y_{am}]$}

\noindent
\resizebox{\textwidth}{!}{\parbox{\textwidth}{%
\begin{align*}
    \E[Y_{am}]
    &=\E\{\E[Y_{am}\mid C]\} & \text{\scriptsize(iterated expectation)}
    \\
    &=\E\{\E[Y_{am}\mid C,A=a]\} & \text{\scriptsize(conditional independence $A\independent Y_{am}\mid C$)}
    \\
    &=\E(\E\{\E[Y_{am}\mid C,A=a,L]\mid C,A=a\}) & \text{\scriptsize(iterated expectation)}
    \\
    &=\E(\E\{\E[Y_{am}\mid C,A=a,L,M=m]\mid C,A=a\}) & \text{\scriptsize(conditional independence $M\independent Y_{am}\mid C,A=a,L$)}
    \\
    &=\E(\E\{\E[Y\mid C,A=a,L,M=m]\mid C,A=a\}). & \text{\scriptsize(consistency and positivity)}
\end{align*}%
}}

\noindent Using the index notation, this result is stated in the main text as
$$\E[Y_{am}]=\E_C(\E_{L\mid C,A=a}\{\E[Y\mid C,A=a,L,M=m]\}).$$

\bigskip

In the following, for simplicity, the notation we use treats the mediator as a discrete variable. The reasoning is the same in the case where the mediator is not discrete, but requires more complicated notation. 

\subsection*{Identification of $\E[Y_{a\M}]$ where $\M$ is a known distribution}

\noindent\textbf{If the interventional mediator distribution $\M$ is unconditional.}
Denote the probability of mediator value $m$ in the distribution $\M$ by $p_\M(m)$. Since $\M$ is defined to be the same distribution for all units, the probability $p_\M(m)$ applies regardless of what values $C,A,L$, etc. take.

\noindent
\resizebox{\textwidth}{!}{\parbox{\textwidth}{%
\begin{align*}
    \E[Y_{a\M}]
    &=\sum_mp_\M(m)\E[Y_{am}] & \text{\scriptsize(iterated expectation)}
    \\
    &=\sum_mp_\M(m)\E(\E\{\E[Y\mid C,A=a,L,M=m]\mid C,A=a\}) & \text{\scriptsize(plugging in the result for $\E[Y_{am}]$)}
    \\
    &=\E(\E\{\sum_mp_\M(m)\E[Y\mid C,A=a,L,M=m]\mid C,A=a\}) & \text{\scriptsize(linearity of expectation)}
    \\
    &=\E[\E(\E_\M\{\E[Y\mid C,A=a,L,M]\mid C,A=a,L\}\mid C,A=a)]. & \text{\scriptsize(rewriting in short form)}
\end{align*}%
}}

\noindent In the last line $\E_\M\{\cdot\}$ indicates that the expectation is taken with respect to the interventional mediator distribution $\M$ instead of the observed mediator distribution.

\medskip

\noindent\textbf{If $\M$ conditions on $C$ but not $L_a$.}
Denote the probability of mediator value $m$ given $C$ in the distribution $\M$ by $p_\M(m\mid C)$. Since $\M$ is defined to be the same distribution for all units that share the same value of $C$, within levels of $C$, the probability $p_\M(m\mid C)$ applies regardless of what values $A,L$, etc. take.

\noindent
\resizebox{\textwidth}{!}{\parbox{\textwidth}{%
\begin{align*}
    \E[Y_{a\M}]
    &=\E\{\E[Y_{a\M}\mid C]\} & \text{\scriptsize(iterated expectation)}
    \\
    &=\E\left\{\sum_m p_\M(m\mid C)\E[Y_{am}\mid C]\right\} & \text{\scriptsize(iterated expectation)}
    \\
    &=\E\left(\sum_mp_\M(m\mid C)\E\{\E[Y\mid C,A=a,L,M=m]\mid C,A=a\}\right) & \text{\scriptsize(plugging in the result for $\E[Y_{am}\mid C]$)}
    \\
    &=\E\left(\E\left\{\sum_mp_\M(m\mid C)\E[Y\mid C,A=a,L,M=m]\mid C,A=a\right\}\right) & \text{\scriptsize(linearity of expectation)}
    \\
    &=\E[\E(\E_\M\{\E[Y\mid C,A=a,L,M]\mid C,A=a,L\}\mid C,A=a)]. & \text{\scriptsize(rewriting in short form)}
\end{align*}%
}}

\medskip

\noindent\textbf{If $\M$ conditions on $C$ and $L_a$ (or just $L_a$).} 
By the decomposition and weak union rules of conditional probability \citep{Koller2009}, the conditional independence assumption $A\independent(L_a,Y_{am})\mid C$ implies
\begin{align*}
    &A\independent L_a\mid C,~~A\independent Y_{am}\mid C, & \text{\scriptsize(decomposition)}
    \\
    &A\independent Y_{am}\mid C,L_a. & \text{\scriptsize(weak union)}
\end{align*}
% We provide proof for completeness.

% \begin{proof}[Decomposition] By assumption,
% \begin{align*}
%     \P(A,L_a,Y_{am}\mid C)
%     &=\P(A\mid C)\P(L_a,Y_{am}\mid C).
% \end{align*}
% Marginalizing $Y_{am}$ out of both sides obtains
% $$\P(A,L_a\mid C)=\P(A\mid C)\P(L_a\mid C),$$
% that is, $A\independent L_a\mid C$. Similar reasoning gives $A\independent Y_{am}\mid C$.
% \end{proof}
% \begin{proof}[Weak union]
% \begin{align*}
%     \P(A,Y_{am}\mid C,L_a)
%     &=\P(Y_{am}\mid C,L_a)\P(A\mid C,L_a,Y_{am})
%     \\
%     &=\P(Y_{am}\mid C,L_a)\P(A\mid C) & \text{\scriptsize(by assumption)}
%     \\
%     &=\P(Y_{am}\mid C,L_a)\P(A\mid C,L_a), & \text{\scriptsize(decomposition result $A\independent L_a\mid C$)}
% \end{align*}
%     that is, $A\independent Y_{am}\mid C,L_a$.
% \end{proof}

% Getting back to the problem at hand, d
Denote the probability of mediator value $m$ given $C,L_a$ in the distribution $\M$ by $p_\M(m\mid C,L_a)$. Since $\M$ is defined to be the same distribution for all units that share the same $(C,L_a)$ value combination, within levels of $(C,L_a)$, the probability $p_\M(m\mid C,L_a)$ applies regardless of what value $A$ takes.

% In the derivation below, there are several instances where $L_a$ can be replaced with $L$ (under consistency, i.e., $L=L_a$ if $A=a$), but for the most part we will keep with $L_a$, except where we switch to $L$ to make use of the mediator-outcome conditional independence assumption and then switch back. The motivation is to be consistent with definition of $\M$ as conditioning on $L_a$ and not simply on the observed $L$ (which is a mixture of $L_a$ and $L_{a'}$).

\noindent
\resizebox{\textwidth}{!}{\parbox{\textwidth}{%
\begin{align*}
    \E[Y_{a\M}]
    &=\E(\E\{\E[Y_{a\M}\mid C,L_a]\mid C\}) & \text{\scriptsize(iterated expectation)}
    \\
    &=\E\left(\E\left\{\sum_mp_\M(m\mid C,L_a)\E[Y_{am}\mid C,L_a]\mid C\right\}\right) & \text{\scriptsize(iterated expectation)}
    \\
    &=\E\left(\E\left\{\sum_mp_\M(m\mid C,L_a)\E[Y_{am}\mid C,A=a,L_a]\mid C\right\}\right) & \text{\scriptsize($A\independent Y_{am}\mid C,L_a$)}
    \\
    &=\E\left\{\sum_l\P(L_a=l\mid C)\sum_mp_\M(m\mid C,L_a=l)\E[Y_{am}\mid C,A=a,L_a=l]\right\} & \text{\scriptsize(writing in long form)}
    \\
    &=\E\left\{\sum_l\P(L_a=l\mid C)\sum_mp_\M(m\mid C,L_a=l)\E[Y_{am}\mid C,A=a,L=l]\right\} & \text{\scriptsize(consistency)}
    \\
    &=\E\left\{\sum_l\P(L_a=l\mid C)\sum_mp_\M(m\mid C,L_a=l)\E[Y_{am}\mid C,A=a,L=l,M=m]\right\} & \text{\scriptsize($M\independent Y_{am}\mid C,A=a,L$)}
    \\
    &=\E\left\{\sum_l\P(L_a=l\mid C)\sum_mp_\M(m\mid C,L_a=l)\E[Y_{am}\mid C,A=a,L_a=l,M=m]\right\} & \text{\scriptsize(consistency)}
    \\
    &=\E\left\{\sum_l\P(L_a=l\mid C)\sum_mp_\M(m\mid C,L_a=l)\E[Y\mid C,A=a,L_a=l,M=m]\right\} & \text{\scriptsize(consistency and positivity)}
    \\
    &=\E\left\{\sum_l\P(L_a=l\mid C,A=a)\sum_mp_\M(m\mid C,L_a=l)\E[Y\mid C,A=a,L_a=l,M=m]\right\} & \text{\scriptsize($A\independent L_a\mid C$)}
    \\
    &=\E\left(\E\left\{\sum_mp_\M(m\mid C,L_a)\E[Y\mid C,A=a,L_a,M=m]\mid C,A=a\right\}\right) & \text{\scriptsize(writing in short form)}
    \\
    &=\E[\E(\E_\M\{\E[Y\mid C,A=a,L_a,M]\mid C,A=a,L_a\}\mid C,A=a)], & \text{\scriptsize(writing in short form)}
    \\
    &=\E[\E(\E_\M\{\E[Y\mid C,A=a,L,M]\mid C,A=a,L\}\mid C,A=a)]. & \text{\scriptsize(consistency)}
\end{align*}%
}}

% \noindent Identification is thus achieved because in the above result, $L_a$ is considered only when $A=a$, which is when it is equal to the observed $L$. In the main text, we replace $L_a$ with $L$ for simplicity. With slight abuse of notation, we replace $L_a$ with $L$ and write
% $$\E[Y_{a\M}]=\E[\E(\E_\M\{\E[Y\mid C,A=a,L,M]\mid C,A=a,L\}\mid C,A=a)].$$

~

\medskip

To sum up, in all three cases of $\M$, the result has the same form in the last line above. Of course, what varies is the definition of $\M$. Using the index notation, we write this in the main text as
$$\E[Y_{a\M}]=\E_C[\E_{L\mid C,A=a}(\E_\M\{\E[Y\mid C,A=a,L,M]\})].$$

\subsection*{Identification of $\E[Y_{a\M}]$ where $\M$ is defined based on a distribution of potential mediator $M_{a^*}$}

\noindent This case inherits the result above, so we only show derivation of distributions of $M_{a^*}$.

\medskip

\noindent First, the distribution conditional on $C$.
\begin{align*}
    \P(M_{a^*}=m\mid C)
    &=\P(M_{a^*}=m\mid C,A=a^*) & \text{\scriptsize(conditional independence $A\independent M_{a^*}\mid C$)}
    \\
    &=\P(M=m\mid C,A=a^*). & \text{\scriptsize(consistency and positivity)}
\end{align*}

\noindent Second, the marginal distribution.
\begin{align*}
    \P(M_{a^*}=m)
    &=\E_C[\P(M_{a^*}=m\mid C)] & \text{\scriptsize(iterated expectation)}
    \\
    &=\E_C[\P(M=m\mid C,A=a^*)]. & \text{\scriptsize(plugging in result for $\P(M_{a^*}=m\mid C)$)}
\end{align*}

\noindent Third, the distribution conditional on $C,L_{a^*}$. By the weak union rule of conditional independence, the assumption $A\independent(L_{a^*},M_{a^*})\mid C$ implies that $A\independent M_{a^*}\mid C,L_{a^*}$.
\begin{align*}
    \P(M_{a^*}=m\mid C,L_{a^*}=l)
    &=\P(M_{a^*}=m\mid C,L_{a^*}=l,A=a^*) & \text{\scriptsize($A\independent M_{a^*}\mid C,L_{a^*}$)}
    \\
    &=\P(M=m\mid C,L=l,A=a^*). & \text{\scriptsize(consistency and positivity)}
\end{align*}

\subsection*{Identification of $\E[Y_{aM_{a'}}]$}

\noindent
\resizebox{\textwidth}{!}{\parbox{\textwidth}{%
\begin{align*}
    \E[Y_{aM_{a'}}]
    &=\E(\E\{\E[Y_{aM_{a'}}\mid M_{a'},C]\mid C\}) & \text{\scriptsize(iterated expectation)}
    \\
    &=\E\left(\sum_m\P(M_{a'}=m\mid C)\E[Y_{aM_{a'}}\mid M_{a'}=m,C]\right) & \text{\scriptsize(writing middle expectation out by definition)}
    \\
    &=\E\left(\sum_m\P(M_{a'}=m\mid C)\E[Y_{am}\mid M_{a'}=m,C]\right) & \text{\scriptsize(consistency)}
    \\
    &=\E\left(\sum_m\P(M_{a'}=m\mid C)\E[Y_{am}\mid C]\right) & \text{\scriptsize(conditional independence $M_{a'}\independent Y_{am}\mid C$)}
    \\
    &=\E\left(\sum_m\P(M_{a'}=m\mid C)\E[Y_{am}\mid C,A=a]\right) & \text{\scriptsize(conditional independence $A\independent Y_{am}\mid C$)}
    \\
    &=\E\left(\sum_m\P(M_{a'}=m\mid C)\E[Y_{am}\mid C,A=a,M=m]\right) & \text{\scriptsize(conditional independence $M\independent Y_{am}\mid C,A=a$)}
    \\
    &=\E\left(\sum_m\P(M_{a'}=m\mid C)\E[Y\mid C,A=a,M=m]\right) & \text{\scriptsize(consistency and positivity)}
    \\
    &=\E\left(\sum_m\P(M_{a'}=m\mid C,A=a')\E[Y\mid C,A=a,M=m]\right) & \text{\scriptsize(conditional independence $A\independent M_{a'}\mid C$)}
    \\
    &=\E\left(\sum_m\P(M=m\mid C,A=a')\E[Y\mid C,A=a,M=m]\right) & \text{\scriptsize(consistency and positivity)}
    \\
    &=\E(\E\{\E[Y\mid C,A=a,M]\mid C,A=a'\}). & \text{\scriptsize(rewriting in short form)}
\end{align*}%
}}

Using the index notation, this result is stated in the main text as
$$\E[Y_{aM_{a'}}]=\E_C(\E_{M\mid C,A=a'}\{\E[Y\mid C,A=a,M]\}).$$

\end{document}